# Harmonic balance-automatic differentiation method: an out-of-the-box and efficient solver for general nonlinear dynamics simulation★

Yi Chen[a,1], Yuhong Jin[b,1], Rongzhou Lin[c], Yifan Jiang[b], Xutao Mei[a,*], Lei Hou[b,*], Yilong Wang[b], Ng Teng Yong[d], Anxin Guo[a]

a. School of Civil Engineering, Harbin Institute of Technology, Harbin 150090, China.

b. School of Astronautics, Harbin Institute of Technology, Harbin 150001, China.

c. Shanghai Institute of Satellite Engineering, Shanghai, 201109, China.

d. School of Mechanical and Aerospace Engineering, Nanyang Technological University, 50 Nanyang Avenue, Singapore 639798, Singapore.

## Abstract

The Harmonic Balance-Alternating Frequency-Time domain (HB-AFT) method is extensively employed for dynamic response analysis of nonlinear systems. However, its application to high-dimensional complex systems is constrained by the manual derivation of Jacobian matrices during Newton-Raphson iterations, which become computationally intractable or error-prone for intricate nonlinearities. The Harmonic Balance-Automatic Differentiation (HB-AD) method is proposed to address this limitation, in which AD is integrated with the harmonic balance framework. This approach eliminates all manual derivations by leveraging AD to compute exact Jacobians numerically, enabling generic and efficient analysis of high-dimensional complex nonlinear systems. The implementation utilizes advanced deep learning frameworks for native parallel computing and CUDA acceleration, and combines AD with arc-length continuation, establishing an out-of-the-box and high efficiency computational architecture. Users need only supply the system's dynamic equations, HB-AD then autonomously trace the complete panorama of periodic responses — including stable/unstable solution branches. Computational experiments on a rotor system with squeeze-film damper (SFD) demonstrate HB-AD's capability in handling complex nonlinear expressions with automated Jacobian calculations. For a high-dimensional aero-engine rotor-bearing-casing system with complex bearing nonlinearities, HB-AD achieves 17-fold higher efficiency than traditional HB-AFT and 144-fold acceleration over the Newmark method. The HB-AD method is a synergistic merger of computational mechanics and machine learning primitives, delivers an easy

★ Source codes are available at https://github.com/shuizidesu/hb-ad.

1 These authors contributed equally to this work.

* Corresponding author

houlei@hit.edu.cn (Lei Hou); meixutao@hit.edu.cn (Xutao Mei)

to use, general-purpose, high efficiency platform for high-fidelity dynamic characterization of high-dimensional engineering systems.

## 1 Introduction

The motion of dynamical systems is typically governed by second-order ordinary differential equations. In practical engineering applications, the primary focus often centers on the long-term steady-state dynamic behavior of the system. Consequently, solving the governing equations of motion to ascertain the steady-state periodic response holds significant engineering importance. The principal solution methodologies encompass three broad categories: numerical, analytical, and semi-analytical techniques. However, to accurately characterize the dynamic behavior of a system, realistic models necessitate not only a high degree of freedom but also the incorporation of nonlinear factors, such as nonlinear stiffness, nonlinear damping, and parametric excitations. These complexities render the determination of the steady-state periodic response for dynamical systems highly challenging[1]. Currently, a robust, easy-to-use, and computationally efficient solution framework remains lacking. Therefore, developing methodologies for solving the equations of motion in nonlinear dynamical systems that concurrently exhibit user-friendliness, broad applicability, and computational efficiency represents substantial theoretical significance and considerable potential for widespread application.

Numerical methods are extensively employed in engineering and scientific computations due to their robust generality and applicability across diverse dynamical systems[2–5]. These techniques are broadly categorized into explicit and implicit approaches, each exhibiting distinct characteristics that dictate their suitability for specific problem types. Explicit methods, such as the Forward Euler method[6], Runge-Kutta schemes (e.g., fourth-order Runge-Kutta (RK4))[7], explicit linear multistep methods[8], derive solutions at each time step solely from known values of preceding steps, thereby offering high computational efficiency by eliminating the need for iterative equation solving and reducing per-step operational costs. This inherent simplicity facilitates straightforward implementation, making them particularly effective for non-stiff system, short-duration simulations, or scenarios with moderate accuracy requirements. For instance, in analyzing nonlinear vibrations of aero-engine rotor systems involving bearing nonlinearities[9,10], SFD nonlinearities[11,12], fault-induced perturbations (e.g., rotor-stator rub[13,14], rotor cracks[15,16], misalignment[17,18], and bearing fault[19,20]), explicit Runge-Kutta methods have proven instrumental due to their computational economy. In contrast, implicit methods—including the Backward Euler method[21], implicit linear multistep methods[22], implicit Runge-Kutta formulations[23], and the Newmark method[24,25] prevalent in structural dynamics—require iterative solutions of nonlinear equation systems within each time step. While this iterative process increases computational overhead, it confers superior numerical stability, notably unconditional stability, and enhanced convergence properties, which are critical for handling stiff systems, long-term simulations, and high-precision applications. Consequently, implicit methods like Newmark are indispensable for high-fidelity dynamic analysis of complex rotor systems characterized by high degrees of freedom and pronounced stiffness, such as those intricate nonlinear effects (e.g. bearing nonlinearities[26,27], damper nonlinearities[28,29], diverse fault nonlinearities[30,31]) or subject to maneuvering flight conditions[32,33]. Despite these advantages, both paradigms exhibit inherent limitations: Explicit methods suffer from conditional stability constraints, rendering them ineffective for stiff equations or high-dimensional nonlinear systems where stringent time-step requirements severely degrade efficiency or induce numerical divergence. Meanwhile, implicit methods, despite their stability, face challenges in computational scalability due to the iterative solution of nonlinear systems at each step, which impedes their utility in large-scale parametric studies of complex models. Furthermore, neither approach can comprehensively capture the full solution topology, as numerical results yield discrete, isolated points that obscure unstable periodic solutions and bifurcation behaviors essential for a holistic understanding of nonlinear dynamics.

Analytical methods[34–36] offer the fundamental advantage of deriving closed-form solutions for the periodic responses of nonlinear dynamical systems, thereby establishing a rigorous theoretical foundation for elucidating intrinsic nonlinear mechanisms while exhibiting high computational efficiency. These methods enable the precise identification of unstable periodic solutions[37] comprehensively characterize the topology of solution branches, and provide profound insights into complex dynamic phenomena including primary resonance[38,39], combination resonance[40,41], internal resonance[42,43], harmonic resonance[44], multistability [45], bifurcation characteristics [46], and instability mechanisms[47]. Consequently, they have been extensively adopted in disciplines such as flight mechanics, fluid dynamics, and structural vibrations. Common analytical approaches encompass perturbation methods[48,49], multiple scale methods[50,51], and the harmonic balance method (HBM)[52,53]. Perturbation and multiple scale methods rely intrinsically on small-parameter expansions[54,55], limiting their applicability predominantly to weakly nonlinear systems. Although strategies such as non-trigonometric basis functions (e.g., elliptic functions), parameter transformations, or multi-scale couplingcan partially extend their scope, these adaptations significantly escalate formulation complexity and compromise algorithmic portability due to their dependence on manual derivations and the absence of generalized computational frameworks. Consequently, these methods remain largely confined to low-degree-of-freedom systems and prove inadequate for high-dimensional nonlinear problems. In contrast, HBM[56,57] constructs steady-state responses through a finite superposition of harmonic basis functions (e.g., Fourier series), transforming differential equations into algebraic systems via coefficient balancing. This approach demonstrates streamlined derivations, robust adaptability to high-dimensional systems, and efficacy in addressing strongly nonlinear regimes[58,59]. Nevertheless, HBM exhibits critical limitations: it fails to handle non-polynomial nonlinearities (e.g., exponential or logarithmic terms) due to the intractability of harmonic expansions, and its practical utility is severely constrained in systems with multi-frequency coupled responses where manual derivation of high-order harmonic balance equations incurs prohibitive computational costs. These inherent constraints in both paradigms underscore a persistent challenge in achieving a unified framework for efficiently resolving high-dimensional nonlinear dynamics with rigorous accuracy and broad applicability.

Semi-analytical methods[60,61], as an integrated paradigm combining analytical and numerical approaches, harness the core advantages of both frameworks. Compared to conventional numerical techniques, they achieve superior computational expediency in capturing long-term steady-state periodic responses[62,63] while comprehensively resolving all solution branches—including unstable periodic orbits[64,65]. This capability enables full reconstruction of the solution-space topology for dynamical systems, thereby establishing a theoretical foundation for probing bifurcation mechanisms[66,67] and instability phenomena[68,69]. Relative to purely analytical approaches, semi-analytical methods offer streamlined formulations and enhanced algorithmic portability, facilitating the development of general-purpose nonlinear solvers. Their robust applicability extends to high-dimensional strongly nonlinear regimes, maintaining efficacy even for systems with non-polynomial nonlinearities (e.g., transcendental-logarithmic couplings or discontinuous piecewise terms)[70,71]. Consequently, these methods have been extensively deployed in aerospace mechanisms[72,73], mechanical systems[74,75], and structural vibrations[76,77]. The Harmonic Balance Alternating Frequency/Time-domain method (HB-AFT)[78] and Incremental Harmonic Balance method (IHB)[79] epitomize this paradigm. Both operate within a harmonic balance-Newton iteration framework[80], wherein the core methodology approximates periodic responses via truncated Fourier series, transforms governing equations into algebraic systems through harmonic balancing, and solves them via Newton iteration. IHB originated from Lau and Cheung’s 1981 seminal work[81], while HB-AFT—integrating harmonic balancing with alternating frequency /time-domain—was formalized by Kim and Noah in 1989[82]. Evolutionary advancements over decades have endowed these methods with exceptional computational performance: Arc-length continuation[83–85] enables exhaustive tracing of solution branches encompassing unstable states; Fast Fourier Transform (FFT)[86–88] and tensor contraction[89,90] synergistically accelerate Jacobian matrix computations; dimensionality reduction[91–93] confines Newton iterations to nonlinear degrees of freedom, slashing computational overhead; and harmonic selection algorithms[94–96] optimize frequency component inclusion, enhancing solution precision. Notwithstanding these merits, a critical bottleneck persists: Jacobian matrix computation remains tethered to symbolic or numerical differentiation paradigms. This process necessitates labor-intensive manual derivations, becoming prohibitively cumbersome even impossible for high-dimensional systems with heterogeneous nonlinear couplings (e.g., concurrent exponential growth, gap constraints, and hysteresis effects). Such limitations impose operational scalability barriers, curtailing broader engineering adoption.

In recent years, transformative progress has been observed in the development of generic methodologies for determining the periodic responses of nonlinear systems.

The latest version of the ubiquitous numerical solver software ANSYS[97] now incorporates the HB-AFT method, thereby furnishing preliminary capabilities for analyzing the dynamics of nonlinear systems. This advancement offers a potent instrument for characterizing the nonlinear behavior inherent to complex engineering systems. However, a significant bottleneck persists in the current implementation: its applicability remains constrained, particularly to systems exhibiting polynomial-type nonlinearities or contact problems. Critically, this approach proves inadequate for effectively addressing engineering systems governed by intricate nonlinear coupling mechanisms, such as those involving compound effects comprising fractional exponential function. Malte Krack and Johann Gross pioneered NLvib—a dedicated nonlinear vibration analysis program[98,99]. This tool synergizes HB-AFT, shooting methods, and arc-length continuation techniques, leveraging MATLAB's fsolve function to automate Jacobian matrix computations and significantly enhance solving usability. Nevertheless, NLvib still confronts dual challenges: the fsolve-dependent Jacobian calculation inevitably accumulates truncation and round-off errors, impairing precision in characterizing multi-solution branches, bifurcation behaviors, and instability mechanisms triggered by strong nonlinear resonances; concurrently, its validation cases are restricted to low-dimensional Duffing systems and rudimentary contact/friction models, while the MATLAB-coded architecture lacks out-of-the-box engineering-ready packaging, thereby hindering direct deployment for dynamic analysis and optimization of high-dimensional complex nonlinear systems in practical engineering context.

This paper proposes a Harmonic Balance with Automatic Differentiation (HB-AD) framework to enable universal efficient analysis of high-dimensional complex nonlinear systems by integrating Automatic Differentiation (AD) into the harmonic balance formulation. The HB-AD method fundamentally eliminates the need for manual derivation of Jacobian matrices—overcoming the inherent limitations of symbolic differentiation (expression explosion) and numerical differentiation (truncation/round-off errors)—while significantly enhancing generalizability and usability for complex nonlinearities. By coupling AD with arc-length continuation and harnessing GPU parallelism (CUDA acceleration) via deep learning infrastructure, HB-AD establishes an out-of-box computational architecture that autonomously resolves complete periodic responses, including both stable and unstable solution branches, directly from user-supplied dynamic equations. **Table. 1** summarizes the comparative advantages and limitations of various methods for obtaining periodic responses in nonlinear systems, revealing that the proposed HB-AD method exhibits high generality and efficiency, with out-of-the-box usability, demonstrating direct applicability for efficient analysis of dynamic characteristics in complex nonlinear systems within practical engineering context.

**Table. 1** A summary of the performance of different methods for obtaining periodic response of nonlinear systems, including the HB-AD method proposed in this paper.

| Methodology | Generality | Usability | Efficiency |
|---|---|---|---|
| HB-AFT | ✔ | ✘ | ✘ |
| DRF-HB [93] | ✔ | ✘ | ✔ |
| ANSYS [97] | ✘ | ✔ | ✘ |
| NLvib [100] | ✔ | ✘ | ✔ |
| HB-AD (ours) | ✔ | ✔ | ✔ |

The subsequent sections of this paper are structured as follows: Section 2 details the fundamental principles and implementation procedures of the proposed HB-AD method. Section 3 demonstrates the superior computational performance of HB-AD through dynamic response analysis of a rotor system with SFD and an aero-engine rotor-bearing-casing system. Section 4 provides further discussion on the specific capabilities and performance metrics of the HB-AD framework. The paper concludes with key findings and contributions summarized in Section 5.

# 2 Methodology

This section proposes the HB-AD (Harmonic balance-automatic differentiation) method. First, the HB-AFT (Harmonic balance-alternating frequency/time domain) method is briefly introduced. Subsequently, by integrating automatic differentiation (AD) techniques into the harmonic balance (HB) procedure, the HB-AD method is formulated. Its fundamental workflow and theoretical principles are comprehensively elaborated.

## 2.1 HB-AFT method

The general differential equations of motion for a $n$-degree-of-freedom ($n$-DOF) nonlinear system can be expressed as follows:

$$\mathbf{M}\ddot{\mathbf{X}}+\mathbf{C}\dot{\mathbf{X}}+\mathbf{K}\mathbf{X}+\mathbf{f}_{\mathrm{N}}(\ddot{\mathbf{X}},\dot{\mathbf{X}},\mathbf{X},t)=\mathbf{F}(t), \tag{1}$$

where **M**, **C**, and **K** represent the mass, damping, and stiffness matrices of the system, respectively; **X** denotes the displacement response vector; $\dot{\mathbf{X}}$ and $\ddot{\mathbf{X}}$ are the first and second derivatives of **X** with respect to time *t*, corresponding to the velocity and acceleration responses. The term $\mathbf{f}_{\mathrm{N}}(\ddot{\mathbf{X}},\dot{\mathbf{X}},\mathbf{X},t)$ encapsulates nonlinearities, which may be functions of displacement **X**, velocity $\dot{\mathbf{X}}$, acceleration $\ddot{\mathbf{X}}$, time *t*, or their coupled interactions. **F**(*t*) represents the external excitation under forced vibration. If **F**(*t*) is periodic excitation with frequency $\omega$, introducing the dimensionless time variable $\tau=\omega t$, yields:

$$\dot{\mathbf{X}}=\frac{d\mathbf{X}}{d\tau}\frac{d\tau}{dt}=\omega\mathbf{X}',\ \ddot{\mathbf{X}}=\frac{d\dot{\mathbf{X}}}{d\tau}\frac{d\tau}{dt}=\omega^2\mathbf{X}''. \tag{2}$$

in which, $(\bullet)'$ and $(\bullet)''$ denote the first-order derivative and the second-order derivative with respect to $\tau$, respectively, then the system's dynamic equations can be normalized as follows:

$$\omega^2\mathbf{M}\mathbf{X}''+\omega\mathbf{C}\mathbf{X}'+\mathbf{K}\mathbf{X}+\mathbf{f}_{\mathrm{N}}(\mathbf{X}'',\mathbf{X}',\mathbf{X},\tau)=\mathbf{F}(\tau), \tag{3}$$

Based on Eq. (3), the residual error of the system dynamics equations is defined as:

$$\mathbf{R}(\mathbf{X}'',\mathbf{X}',\mathbf{X},\tau)\triangleq\omega^2\mathbf{M}\mathbf{X}''+\omega\mathbf{C}\mathbf{X}'+\mathbf{K}\mathbf{X}+\mathbf{f}_{\mathrm{N}}(\mathbf{X}'',\mathbf{X}',\mathbf{X},\tau)-\mathbf{F}(\tau), \tag{4}$$

Since the external excitation F(*t*) is periodic, the periodic response **X** and residual **R** of the system can be expressed via truncated Fourier series as:

$$\mathbf{X}=\mathbf{A}\mathbf{H},\ \mathbf{R}=\mathbf{B}\mathbf{H}, \tag{5}$$

where, **H** denotes the harmonic basis matrix of the Fourier series, and **A**, **B** are the Fourier coefficient matrices corresponding to the expansions of **X** and **R**, respectively, which can be explicitly expressed as:

$$\mathbf{H}=\begin{pmatrix}1 & \cos\tau & \cdots & \cos s\tau & \sin\tau & \cdots & \sin s\tau\end{pmatrix}^{\mathrm{T}}, \tag{6}$$

$$\mathbf{A}=\begin{bmatrix} a_{01} & a_{11} & \cdots & a_{s1} & b_{11} & \cdots & b_{s1} \\ \vdots & \vdots & \ddots & \vdots & \vdots & \ddots & \vdots \\ a_{0n} & a_{1n} & \cdots & a_{sn} & b_{1n} & \cdots & b_{sn} \end{bmatrix}, \tag{7}$$

$$\mathbf{B}=\begin{bmatrix} c_{01} & c_{11} & \cdots & c_{s1} & d_{11} & \cdots & d_{s1} \\ \vdots & \vdots & \ddots & \vdots & \vdots & \ddots & \vdots \\ c_{0n} & c_{1n} & \cdots & c_{sn} & d_{1n} & \cdots & d_{sn} \end{bmatrix}, \tag{8}$$

where, *s* denotes the order of the Fourier series expansion, which depends on the nature of the external excitation and nonlinearities. Subsequently, by employing the harmonic balance procedure, the differential equations (Eq. (3)) are transformed into a set of algebraic equations, expressed as:

$$\mathbf{B}(\mathbf{A}) = \mathbf{0}. \tag{9}$$

To obtain the solution of Eq. (9), the Newton-Raphson iterative scheme is introduced as follows:

$$\mathbf{A}^{(k+1)} = \mathbf{A}^{(k)} - \mathbf{J}^{-1}\mathbf{B},\ \mathbf{J} = \partial\mathbf{B}/\partial\mathbf{A}, \tag{10}$$

in which, $\mathbf{J}$ is the Jacobian matrix. Since the nonlinear term $\mathbf{f}_{\mathrm{N}}(\mathbf{X}'', \mathbf{X}', \mathbf{X}, \tau)$ depends on time $\tau$, the response $\mathbf{X}$, and its derivatives $\mathbf{X}'$, $\mathbf{X}''$, the Fourier coefficients of the residual $\mathbf{R}$ consequently exhibit dependencies on these variables. When $\mathbf{f}_{\mathrm{N}}(\mathbf{X}'', \mathbf{X}', \mathbf{X}, \tau)$ assumes complex functional forms, such as fractional exponentials, the Jacobian matrix $\mathbf{J}$ cannot be directly derived. To address this, the alternating frequency/time domain (AFT) method is introduced, with implementation details outlined as follows.

Discrete sampling is first performed in the time domain, specifically: an entire period $T$ ($T$=2π/$\omega$) of the system's responses is uniformly partitioned into $N$ segments, thereby obtaining a discrete time sequence as follows:

$$\tau_k = \frac{T}{N}k,\ k = 0,1,2,\ \cdots,\ N-1. \tag{11}$$

Subsequently, the time-domain expression of the residual Fourier coefficients is obtained via the inverse discrete Fourier transform (IDFT), formulated as:

$$\begin{gathered} \mathrm{c}_{0i} = \frac{1}{N}\sum_{k=0}^{N-1} R_i(X'', X', X, \tau_k),\ \mathrm{c}_{ji} = \frac{2}{N}\sum_{k=0}^{N-1} R_i(X'', X', X, \tau_k)\cos j\tau_k, \\ \mathrm{d}_{ji} = \frac{2}{N}\sum_{k=0}^{N-1} R_i(X'', X', X, \tau_k)\sin j\tau_k,\ (j=1-s,\ i=1-n), \end{gathered} \tag{12}$$

Note that $\mathbf{X} = \mathbf{AH}$, then we can write

$$\mathbf{X}'' = \mathbf{AH}'',\ \mathbf{X}' = \mathbf{AH}' \tag{13}$$

By considering Equations (4), (5), (10), (12), and (13), the Jacobian matrix $\mathbf{J}$ can be derived via the chain rule of differentiation, and expressed as:

$$\mathbf{J} = \frac{\partial\mathbf{B}}{\partial\mathbf{A}} = \begin{bmatrix} \partial\mathbf{B}_1/\partial\mathbf{A}_1 & \cdots & \partial\mathbf{B}_1/\partial\mathbf{A}_i & \cdots & \partial\mathbf{B}_1/\partial\mathbf{A}_n \\ \vdots & \ddots & \vdots & \ddots & \vdots \\ \partial\mathbf{B}_i/\partial\mathbf{A}_1 & \cdots & \partial\mathbf{B}_i/\partial\mathbf{A}_i & \cdots & \partial\mathbf{B}_i/\partial\mathbf{A}_n \\ \vdots & \ddots & \vdots & \ddots & \vdots \\ \partial\mathbf{B}_n/\partial\mathbf{A}_1 & \cdots & \partial\mathbf{B}_n/\partial\mathbf{A}_i & \cdots & \partial\mathbf{B}_n/\partial\mathbf{A}_n \end{bmatrix}, \tag{14}$$

and here, $\mathbf{B}_i$ denotes the $i$-th row of the residual Fourier coefficient matrix, representing the Fourier coefficients of the residual on the $i$-th DOF, while $\mathbf{A}_i$ is the $i$-th row of the response Fourier coefficient matrix, corresponding to the Fourier coefficients of the dynamic response on the same DOF. Thereby, the Jacobian matrix $\mathbf{J}$ can be calculated by the chain derivation rule, i.e.,

$$\begin{aligned}\frac{\partial \mathbf{B}_i}{\partial \mathbf{A}_i}&=\frac{\partial \mathbf{B}_i}{\partial \mathbf{f}_{\mathrm{N}_i}}\left(\frac{\partial \mathbf{f}_{\mathrm{N}_i}}{\partial \mathbf{X}''_i}\frac{\partial \mathbf{X}''_i}{\partial \mathbf{A}_i}+\frac{\partial \mathbf{f}_{\mathrm{N}_i}}{\partial \mathbf{X}'_i}\frac{\partial \mathbf{X}'_i}{\partial \mathbf{A}_i}+\frac{\partial \mathbf{f}_{\mathrm{N}_i}}{\partial \mathbf{X}_i}\frac{\partial \mathbf{X}_i}{\partial \mathbf{A}_i}\right),\\ \frac{\partial \mathbf{B}_i}{\partial \mathbf{A}_j}&=\frac{\partial \mathbf{B}_i}{\partial \mathbf{R}_i}\left(\frac{\partial \mathbf{f}_{\mathrm{N}_i}}{\partial \mathbf{X}''_j}\frac{\partial \mathbf{X}''_j}{\partial \mathbf{A}_j}+\frac{\partial \mathbf{f}_{\mathrm{N}_i}}{\partial \mathbf{X}'_j}\frac{\partial \mathbf{X}'_j}{\partial \mathbf{A}_j}+\frac{\partial \mathbf{f}_{\mathrm{N}_i}}{\partial \mathbf{X}_j}\frac{\partial \mathbf{X}_j}{\partial \mathbf{A}_j}\right),\end{aligned} \tag{15}$$

where, $\mathbf{X}_i$ , $\mathbf{X}_i'$ and $\mathbf{X}_i''$ denote the displacement, velocity and acceleration responses of the system on the $i$-th DOF, respectively. Furthermore, substituting $\mathbf{B}_i=\begin{pmatrix}c_{0i} & c_{1i} & \cdots & c_{si} & d_{1i} & \cdots & d_{si}\end{pmatrix}, \mathbf{A}_i=\begin{pmatrix}a_{0i} & a_{1i} & \cdots & a_{si} & b_{1i} & \cdots & b_{si}\end{pmatrix}$ into Eq. (14), and utilizing the chain derivation rule, the detailed expressions of $\partial \mathbf{B}_i / \partial \mathbf{A}_i$ can be obtained:

$$\frac{\partial \mathbf{B}_i}{\partial \mathbf{A}_i}=\begin{bmatrix}\partial c_{0i}/\partial a_{0i} & \partial c_{0i}/\partial a_{1i} & \cdots & \partial c_{0i}/\partial a_{si} & \partial c_{0i}/\partial b_{1i} & \cdots & \partial c_{0i}/\partial b_{si}\\ \partial c_{1i}/\partial a_{0i} & \partial c_{1i}/\partial a_{1i} & \cdots & \partial c_{1i}/\partial a_{si} & \partial c_{1i}/\partial b_{1i} & \cdots & \partial c_{1i}/\partial b_{si}\\ \vdots & \vdots & \ddots & \vdots & \vdots & \ddots & \vdots\\ \partial c_{si}/\partial a_{0i} & \partial c_{si}/\partial a_{1i} & \cdots & \partial c_{si}/\partial a_{si} & \partial c_{si}/\partial b_{1i} & \cdots & \partial c_{si}/\partial b_{si}\\ \partial d_{1i}/\partial a_{0i} & \partial d_{1i}/\partial a_{1i} & \cdots & \partial d_{1i}/\partial a_{si} & \partial d_{1i}/\partial b_{1i} & \cdots & \partial d_{1i}/\partial b_{si}\\ \vdots & \vdots & \ddots & \vdots & \vdots & \ddots & \vdots\\ \partial d_{si}/\partial a_{0i} & \partial d_{si}/\partial a_{1i} & \cdots & \partial d_{si}/\partial a_{si} & \partial d_{si}/\partial b_{1i} & \cdots & \partial d_{si}/\partial b_{si}\end{bmatrix} \tag{16}$$

in which,

$$\frac{\partial c_{0i}}{\partial a_{ji}}=\frac{1}{N}\sum_{k=0}^{N-1}\left(\left(\frac{\partial \mathbf{f}_{\mathrm{N}_i}}{\partial \mathbf{X}_i}-j^2\frac{\partial \mathbf{f}_{\mathrm{N}_i}}{\partial \mathbf{X}''_i}\right)\cos j\tau_k - j\frac{\partial \mathbf{f}_{\mathrm{N}_i}}{\partial \mathbf{X}'_i}\sin j\tau_k\right) \tag{17}$$

$$\frac{\partial c_{mi}}{\partial a_{ji}}=\frac{2}{N}\sum_{k=0}^{N-1}\left(\left(\left(\frac{\partial \mathbf{f}_{\mathrm{N}_i}}{\partial \mathbf{X}_i}-j^2\frac{\partial \mathbf{f}_{\mathrm{N}_i}}{\partial \mathbf{X}''_i}\right)\cos j\tau_n - j\frac{\partial \mathbf{f}_{\mathrm{N}_i}}{\partial \mathbf{X}'_i}\sin j\tau_k\right)\cos m\tau_k\right) \tag{18}$$

$$\frac{\partial c_{mi}}{\partial b_{ji}}=\frac{2}{N}\sum_{k=0}^{N-1}\left(\left(\left(\frac{\partial \mathbf{f}_{\mathrm{N}_i}}{\partial \mathbf{X}_i}-j^2\frac{\partial \mathbf{f}_{\mathrm{N}_i}}{\partial \mathbf{X}''_i}\right)\sin j\tau_k + j\frac{\partial \mathbf{f}_{\mathrm{N}_i}}{\partial \mathbf{X}'_i}\cos j\tau_k\right)\cos m\tau_k\right) \tag{19}$$

$$\frac{\partial d_{mi}}{\partial a_{ji}}=\frac{2}{N}\sum_{k=0}^{N-1}\left(\left(\left(\frac{\partial \mathbf{f}_{\mathrm{N}_i}}{\partial \mathbf{X}_i}-j^2\frac{\partial \mathbf{f}_{\mathrm{N}_i}}{\partial \mathbf{X}''_i}\right)\cos j\tau_k - j\frac{\partial \mathbf{f}_{\mathrm{N}_i}}{\partial \mathbf{X}'_i}\sin j\tau_k\right)\sin m\tau_k\right) \tag{20}$$

$$\frac{\partial d_{mi}}{\partial b_{ji}}=\frac{2}{N}\sum_{k=0}^{N-1}\left(\left(\left(\frac{\partial \mathbf{f}_{\mathrm{N}_i}}{\partial \mathbf{X}_i}-j^2\frac{\partial \mathbf{f}_{\mathrm{N}_i}}{\partial \mathbf{X}''_i}\right)\sin j\tau_k + j\frac{\partial \mathbf{f}_{\mathrm{N}_i}}{\partial \mathbf{X}'_i}\cos j\tau_k\right)\sin m\tau_k\right) \tag{21}$$

Once the Jacobian matrix **J** is obtained, the Newton iteration defined in Eq. (10) can be executed to iteratively update the value of **A**. The iteration converges when the norm of **B** falls below a specified error tolerance, thereby yielding the periodic response of the system as **X** = **AH**.

## 2.2 HB-AD method

As established in the preceding analysis, computing the Jacobian matrix is critical to the HB-AFT method for determining periodic responses of nonlinear systems. The computational approach for the Jacobian matrix directly governs both the generality and efficiency of the method. In conventional HB-AFT implementations, the Jacobian matrix required for Newton iteration is typically computed via the inverse Fourier transform coupled with the chain rule. However, this process inevitably necessitates extensive symbolic manipulations and manual derivations, which are not only time-consuming and labor-intensive but also prone to errors. To overcome these limitations, the HB-AD method leverages automatic differentiation to enable automatic and efficient computation of the Jacobian matrix, thereby comprehensively enhancing the performance for obtaining periodic response of nonlinear systems.

Automatic differentiation (AD) serves as the core technology for gradient computation in deep learning frameworks, enabling rapid calculation of Jacobian matrices. Reverse-mode AD is particularly suited for scenarios characterized by high-dimensional inputs and low-dimensional outputs. This aligns precisely with the requirement for Jacobian computation in semi-analytical harmonic balance methods for determining periodic responses of nonlinear systems. Specifically, for the residual generating function, inputs encompass system parameters and initial values of response calculations (high-dimensional), while the output residual vector is low-dimensional. Therefore, we adopt reverse-mode AD to compute the Jacobian matrix, which achieves efficient evaluation through a single forward pass to compute function values, followed by a single backward pass to compute gradients. Notably, reverse-mode AD accommodates functions defined by arbitrary code segments and handles cases where functions are not fully differentiable, provided they are differentiable at the targeted evaluation points. This capability ensures exceptional versatility in practical implementations.

The computational workflow of the HB-AD method for determining the periodic response of nonlinear systems is illustrated in **Fig. 1**. The procedure aligns with the conventional HB-AFT framework, comprising three sequential phases:

(i) Set solutions: the periodic response is approximated using a truncated Fourier series.

(ii) Equation transformation: the governing differential equations are transformed into algebraic equations via the harmonic balance method.

(iii) Iterative solution: Newton-Raphson iteration is employed to solve these algebraic equations to obtain the periodic response.

Notably, in the HB-AD method, the required Jacobian matrix is acquired via AD, thereby eliminating the need for intricate manual derivation of analytical expressions. This integration liberates the Jacobian computation from any human-driven symbolic manipulation, significantly enhancing the method's accessibility and enabling out-of-the-box usability.

As illustrated in **Fig. 1**, the HB-AD method provides a principled and streamlined framework for determining periodic responses of nonlinear systems. This clarity in workflow enables intuitive program architecture design, significantly enhancing code readability, portability, and generality. Furthermore, leveraging the extensive optimizations for parallel computing and CUDA integration in advanced deep learning frameworks such as PyTorch and JAX, the proposed HB-AD method directly utilizes automatic differentiation operators (e.g., jacrev) to enable efficient Jacobian matrix computations. By introducing vectorization via vmap during this process, the method eliminates iterative loops for Jacobian calculations, thereby significantly accelerating computational efficiency. Concurrently, HB-AD leverages native GPU parallelization architecture to process large-scale problems with exceptional throughput. This capability proves particularly advantageous in engineering practice when conducting large-scale parametric studies, where its parallelized design elevates computational efficiency by orders of magnitude and substantially compresses analysis timelines.

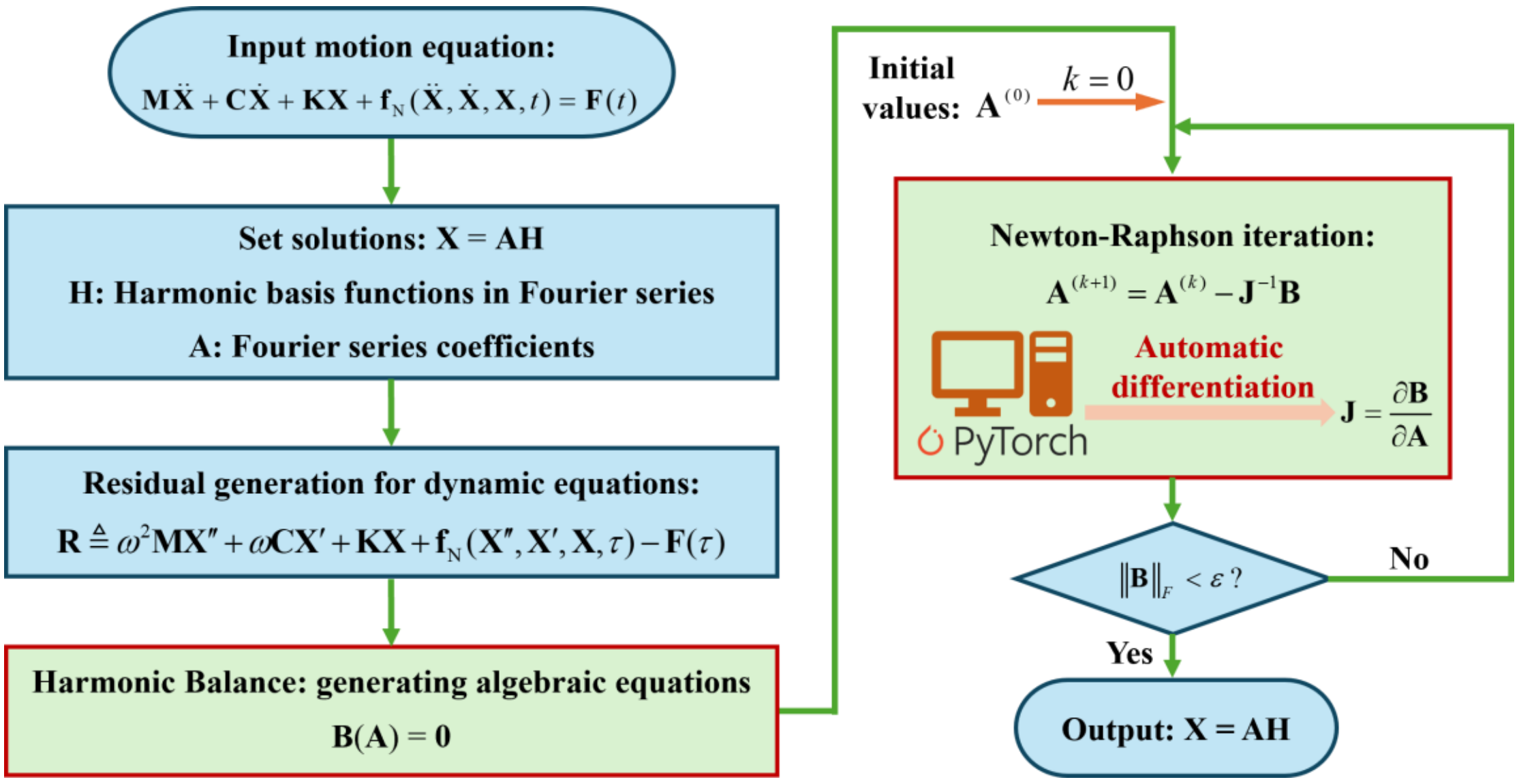

**Fig. 1** Flowchart of the HB-AD method. Acquiring periodic responses of nonlinear systems via HB-AD method comprises two primary phases: (i) transforming differential equations into algebraic equations through the Harmonic balance method (left panel of Fig. 1), (ii) solving the derived algebraic equations via Newton iteration (right panel of Fig. 1). During Newton iteration, automatic differentiation is employed to compute the requisite Jacobian matrix efficiently, thereby significantly enhancing the method's versatility and computational efficiency.

Complex nonlinear systems frequently exhibit multi-solution phenomena. Comprehensively capturing distinct solution branches to characterize periodic responses is paramount for elucidating intrinsic nonlinear dynamic mechanisms, such as bifurcation and instability. While conventional techniques — including the HB method, IHB, and HB-AFT — can integrate arc-length continuation technology[84] to trace solution branches, their implementation faces a critical bottleneck: existing frameworks universally require manual analytical derivationof the partial derivative of the algebraic residual (obtain by HB procedure ) with respect to the continuation parameter (e.g., rotational speed in rotor systems). This dependency on hand-crafted Jacobians renders systematic exploration of solution branches via arc-length continuation cumbersome or even computationally infeasiblefor high-dimensional complex systems.

To overcome this limitation, we synergize AD with arc-length continuation within the HB-AD framework. This integration enables automated computationof the residual's partial derivative—a core component of the Newton iteration's Jacobian matrix—directly from the HB algebraic equations.Notably, AD eliminates the need for error-prone manual derivation, thereby:

(1) Generalizing the continuation processacross arbitrary nonlinearities (e.g., transcendental functions, hysteresis) without symbolic manipulation;
(2) Enhancing computational robustness by avoiding approximation errors inherent in finite-difference schemes;
(3) Streamlining solution-branch tracingfor systems with high-dimensional state variables or complex coupling.

Consequently, HB-AD method significantly elevates usability for high-fidelity analysis of complex nonlinear systems. The framework delivers out-of-the-box usabilityfor reconstructing the complete panorama of periodic responses—encompassing stable/unstable branches and bifurcation points—directly from the nonlinear systems' dynamic equations.

**Listing 1** exemplifies key Python implementations of the HB-AD method, revealing a logically coherent structure. Users need only define the target nonlinear system—specifying the mass matrix **M**, damping matrix **C**, stiffness matrix **K**, external excitation Fex, and nonlinear term nonlinearity—to initiate automated solving. Notably, as shown in Line 27 of Listing 1, the Jacobian matrix required for Newton iteration is automatically computed via reverse-mode automatic differentiation using jacrev, eliminating error-prone manual derivations while ensuring machine-precision accuracy. Collectively, the HB-AD workflow entirely bypasses symbolic manipulations, offering out-of-the-box usability. This approach democratizes access to nonlinear system analysis, empowering practitioners across disciplines to efficiently tackle complex dynamics. Entire project codes and more details about the HB-AD method are provided in https://github.com/shuizidesu/hb-ad.

**Listing. 1** Partial core Python implementation of the HB-AD method for solving periodic responses of nonlinear systems. Users define the system, and the HB-AD method automates the solution process, leveraging automatic differentiation (AD) for Jacobian computation without manual derivation.

```
'''Import the required libraries'''
import torch
from torch.func import jacrev
...
'''Config and time-invariant parameter'''
...
'''Define the system equation'''
class MyNonlinearSystym(torch.nn.Module):
    def _ _init_ _(self):
    ...
    def initialize parameters(self,parameters):
    ...
    def force(self):
    ...
    def nonlinearity(self, x,dx):
    ...
    def calculate_residual(self, hb_coefficient):
    ...
        residual_fft = (torch.fft.rfft(self.omega1**2*self.M @ ddx +
        self.omega1*self.C @ dx + self.K @ x - self.nonlinearity(x,dx) - Fex,
        dim=1)* 2 / num_time_steps)
        return residual_vector
'''Perform Newton-Raphson iteration to compute periodic responses'''
...
# Computing the Jacobi matrix using automatic differentiation reverse mode)through
the jacrev function
jacobian = jacrev(MyNonlinearSystym.calculate_residual)(hb_coefficient)
# Computing the residual vector
residual_equation = MyNonlinearSystym.calculate_residual(hb_coefficient)
# computing the increment
delta_hb_coeff = torch.linalg.solve(jacobian, residual_equation)
# Updating the solution
hb_coeff = hb_coeff - delta_hb_coeff
...
```

In summary, the HB-AD method integrates harmonic balance (HB), Newton-Raphson iteration, and automatic differentiation (AD) into a unified computational framework. The governing differential equations are transformed into algebraic equations via HB. Newton iteration then solves these algebraic equations to determine the system's periodic response. Notably, AD automates the computation of Jacobian matrices required for Newton iteration, eliminating manual symbolic derivations while ensuring machine-precision accuracy. This synergy significantly enhances computational efficiency and universality, enabling an out-of-the-box workflow that requires no user intervention for Jacobian derivation. Consequently, HB-AD establishes a robust foundation for generalized, programmable implementations of nonlinear dynamic analysis.

# 3 Numerical examples

To demonstrate the superior capability of the HB-AD method in solving nonlinear differential equations, this section presents two representative case studies. The first case involves a rotor system with squeeze film damper (SFD), exhibiting complex damping and stiffness nonlinearities. The second case models an aero-engine rotor-bearing-casing system incorporating multiple nonlinearities, including fractional exponential functions, clearances, and time-varying stiffness excitation, under dual-frequency external forcing. All numerical simulations were conducted on a workstation equipped with an Intel(R) Core™ i7-14700KF processor, 64 GB RAM, an NVIDIA GeForce RTX 4070 SUPER GPU, and Windows 10 Pro OS. The computational performance of the HB-AD method is systematically benchmarked against the HB-AFT method, the fourth-order Runge-Kutta (RK4) method, and the Newmark method.

## 3.1 Rotor system with squeeze film damper (SFD)

### 3.1.1 Dynamic model of the rotor system

Consider a rotor system with SFD[101], as depicted in **Fig. 2**. The rotor is supported at both ends by linear elastic bearings, with the SFD installed specifically at the left support. The rotor mass is concentrated at an offset disk.

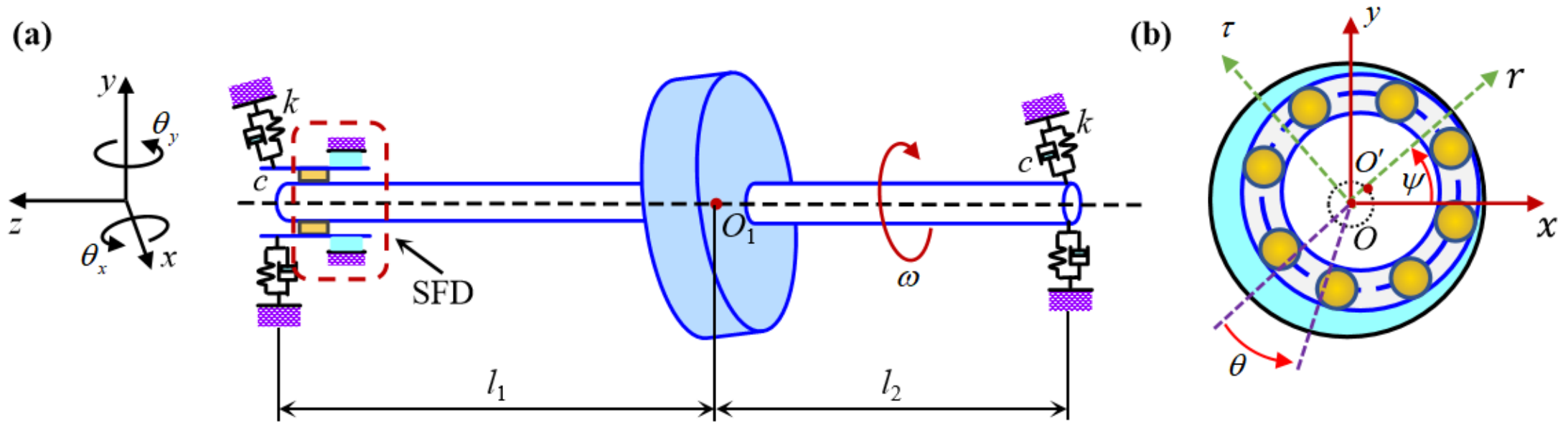

**Fig. 2** Schematic diagram of the rotor system with squeeze film damper. (a) Schematic diagram of the rotor system, (b) Schematic diagram of the SFD. The rotor is supported at both ends by linear elastic bearings, with the SFD installed specifically at the left support. The motion of the rotor is described by four degrees of freedom: namely the displacements in the horizontal and vertical directions, denoted by $x$ and $y$, and the rotations about the horizontal and vertical axes, denoted by $\theta_x$ and $\theta_y$, respectively.

Accounting for both the transverse vibration and rigid-body rotation of the rotor,

by employing the Lagrange's equations of the second kind, the system's nonlinear dynamics equations are derived as follows:

$$\begin{cases} m\ddot{x}+2c\dot{x}+c(l_1-l_2)\dot{\theta}_y+kx+k(l_1-l_2)\theta_y+F_x=me\omega^2\cos(\omega t), \\ m\ddot{y}+2c\dot{y}+c(l_2-l_1)\dot{\theta}_x+ky+k(l_2-l_1)\theta_x+F_y=me\omega^2\sin(\omega t), \\ J_d\ddot{\theta}_x+c(l_2-l_1)\dot{y}+c(l_1^2+l_2^2)\dot{\theta}_x+J_p\omega\dot{\theta}_y+(l_2-l_1)y+k(l_1^2+l_2^2)\theta_x-F_yl_1=0, \\ J_d\ddot{\theta}_y+c(l_1-l_2)\dot{x}+c(l_1^2+l_2^2)\dot{\theta}_y-J_p\omega\dot{\theta}_x+k(l_1-l_2)x+k(l_1^2+l_2^2)\theta_y+F_xl_1=0, \end{cases} \tag{22}$$

where, $m$ denotes the rotor mass, $c$ represents the damping coefficient at the supports, and $k$ designates the support stiffness. The rotor's rotational inertia is characterized by $J_d$ (diametral moment of inertia) and $J_p$ (polar moment of inertia). The geometric configuration is defined by $l_1$ and $l_2$, corresponding to the shaft segment lengths from the disk to the left support and right support, respectively. The rotor imbalance is quantified by the eccentricity $e$, while $\omega$ indicates the rotational speed. The lateral displacement responses in the horizontal ($x$) and vertical ($y$) directions are denoted by $x$ and $y$, with their first-time derivative $\dot{x}$ and $\dot{y}$ representing the lateral velocities, and second-time derivatives $\ddot{x}$ and $\ddot{y}$ indicating the lateral accelerations. The angular displacements about the $x$- and $y$-axes are expressed as $\theta_x$ and $\theta_y$, while $\dot{\theta}_x$ and $\dot{\theta}_y$ signify the angular velocities, and $\ddot{\theta}_x$ and $\ddot{\theta}_y$ represent the angular accelerations. Finally, $F_x$ and $F_y$ denote the nonlinear hydrodynamic force components in the $x$- and $y$-directions generated by the SFD.

For the SFD cross-section illustrated in **Fig. 2**(b), the journal center $O'$ whirls about the geometric center $O$ of the outer housing. Assuming the lubricant is an incompressible fluid and neglecting inertial effects, the Reynolds equation governing the hydrodynamic pressure distribution $p$ along the axial direction $z$ and circumferential coordinate $\theta$ is expressed as:

$$\frac{\partial}{\partial\theta}\left(h^3\frac{\partial p}{\partial\theta}\right)+R^2\frac{\partial}{\partial z}\left(h^3\frac{\partial p}{\partial z}\right)=12\mu R^2(\delta_e\omega\sin\theta+\dot{\delta}_e\cos\theta), \tag{23}$$

where, the film thickness $h$ is defined as $h=\delta_c+\delta_e\cos\theta$, $\delta_c$ denotes the radial clearance of the oil film, $\mu$ denotes the dynamic viscosity of the lubricant, $R$ represents the journal radius, $\delta_e$ signifies the radial displacement of the journal center, expressed as follow:

$$\delta_e=\sqrt{(x+\theta_yl_1)^2+(y-\theta_xl_1)^2}, \tag{24}$$

By defining the circumferential coordinate $\theta = 0$ at the position of maximum film thickness, according to the short bearing assumption, the oil film forces are derived as follows:

$$\begin{cases} F_r = \dfrac{\mu R L^3}{\delta_c^{\ 2}}\left(I_3^{11}\dot{\psi}\delta_r + I_3^{02}\dot{\delta}_r\right), \\ F_t = \dfrac{\mu R L^3}{\delta_c^{\ 2}}\left(I_3^{20}\dot{\psi}\delta_r + I_3^{11}\dot{\delta}_r\right), \end{cases} \tag{25}$$

where, $F_r$ denotes the radial oil film force, $F_T$ denotes the tangential oil film force, $\delta_r = \delta_e / \delta_c$ denotes the dimensionless radial displacement of the journal center, and $\psi$ represents the precession angle of the journal, expressed as:

$$\psi = \arctan\frac{y - \theta_x l_1}{x + \theta_y l_1}, \tag{26}$$

$I$ represents the Sommerfeld integral constant, which can be computed as follows:

$$I_3^{ln} = \int_{\theta_1}^{\theta_2} \frac{\sin^l\theta\cos^n\theta}{(1+\delta_r\cos\theta)^3} d\theta (l, n = 0 \sim 2), \tag{27}$$

in which, $\theta_1$ and $\theta_2$ denote the angular coordinates delimiting the active pressure zone of the oil film, satisfying the following conditions:

$$\begin{cases} \theta_1 = \arctan\left(-\dfrac{\dot{r}}{r\dot{\psi}}\right), \\ \theta_2 = \theta_1 + \pi. \end{cases} \tag{28}$$

Transforming the hydrodynamic oil film force described by Eq. (25) into the Cartesian coordinate system yields the force components in the $x$- and $y$-directions as follows:

$$\begin{cases} F_x = F_r \dfrac{x + \theta_y l_1}{\delta_e} - F_t \dfrac{y - \theta_x l_1}{\delta_e}, \\ F_y = F_r \dfrac{y - \theta_x l_1}{\delta_e} + F_t \dfrac{x + \theta_y l_1}{\delta_e}. \end{cases} \tag{29}$$

Based on Eq. (24) – (29), it can be observed that the oil film force of the SFD exhibits complex nonlinear characteristics. Specifically, it is a nonlinear function of ($x$, $y$, $\theta_x$, $\theta_y$) and their derivatives $(\dot{x}, \dot{y}, \dot{\theta}_x, \dot{\theta}_y)$. These nonlinearities involve fractional power functions, negative exponential functions, inverse trigonometric functions, as well as integral operations. The intricate coupling of these mathematical forms makes it extremely challenging to derive the Jacobian matrix required by Newton's iteration in implicit numerical or analytical methods for computing the system's periodic response. The derivation process is not only cumbersome but also highly error-prone, rendering traditional methods based on Newton's iteration practically infeasible for solving the system's dynamic equations in such cases. In contrast, the HB-AD method proposed in this study demonstrates strong applicability to such complex nonlinear differential equations. Operating within the framework of the Harmonic Balance (HB) method, HB-AD leverages automatic differentiation to compute the Jacobian matrix required for Newton iteration. This approach demonstrates exceptional applicability and robustness for strongly nonlinear systems with intricate functional couplings. Subsequently, the periodic responses of the system are computed using both the HB-AD method and a fourth-order Runge-Kutta (RK4) numerical integration scheme, with comparative analysis elucidating their performance.

### 3.1.2 Results analysis for the rotor system with SFD

The system’s dynamic equation (Eq. (22)) is solved using the HB-AD method and the RK4 method to obtain the periodic responses near the first-order critical speed. Throughout this computational procedure, the Sommerfeld integral, explicitly defined by Eq. (27), was numerically evaluated employing Gaussian-Legendre quadrature. The amplitude-frequency response curve is illustrated in **Fig. 3** (a). Furthermore, point A, selected in the off-resonance region of the amplitude-frequency response, exhibits its time history and rotors’ orbit in **Fig. 3** (b) and (c), respectively. Point B, chosen within the resonance region, displays corresponding results in **Fig. 3** (d) and (e). Notably, the amplitude-frequency responses, time histories, and rotors’ orbits obtained by both methods display excellent agreement, validating the applicability of the HB-AD method for solving complex nonlinear dynamic equations of rotor systems with SFD.

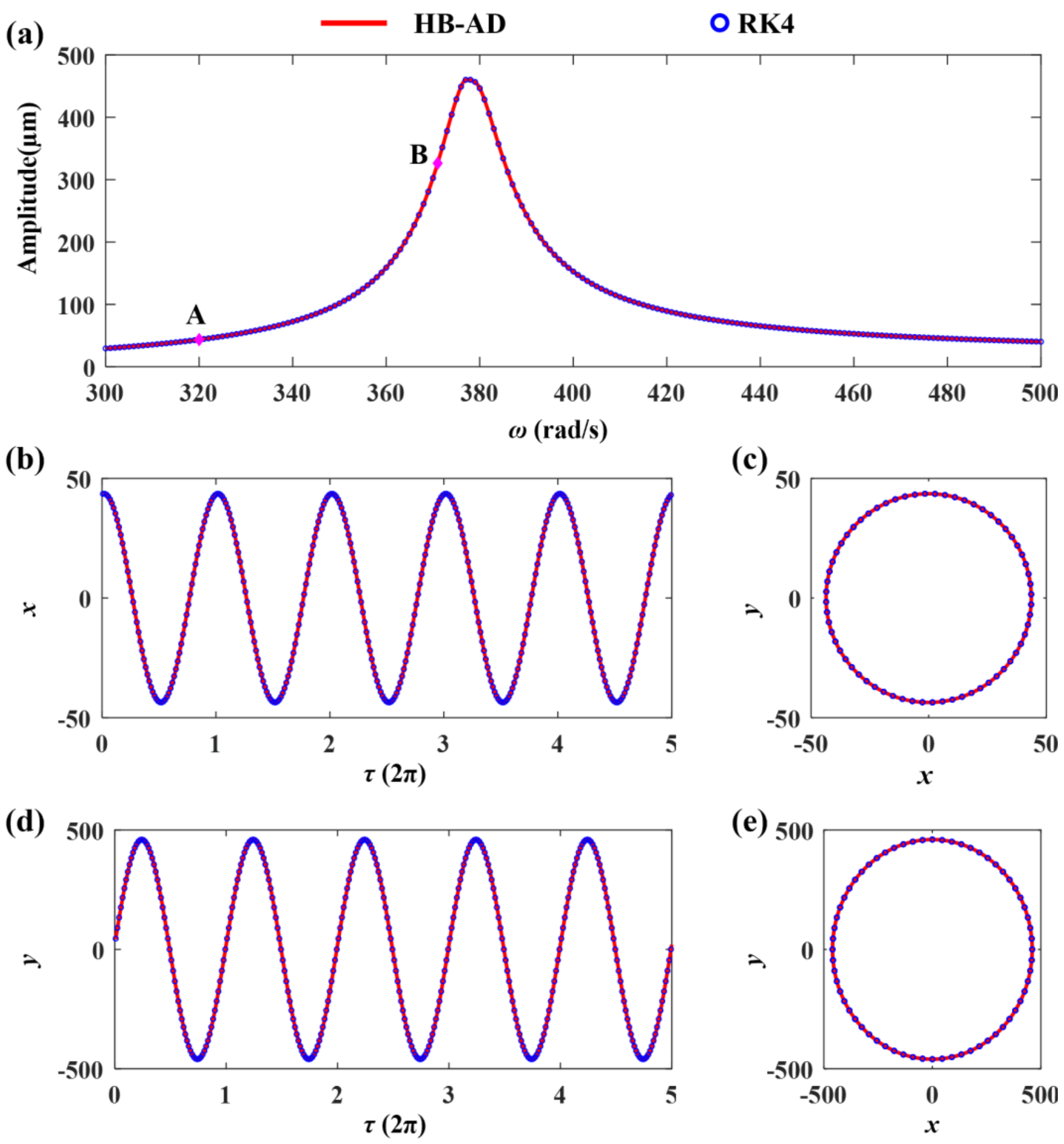

**Fig. 3** Response curves of the rotor systems with SFD. (a) amplitude-frequency response curve near the first-order critical speed, (b) the time history diagram at point A, (c) rotors' orbits at point A, (d) the time history diagram at point B, (c) rotors' orbits at point B. As illustrated, the computational results of the HB-AD method and the RK4 method exhibit excellent agreement.

## 3.2 Aero-engine rotor system with bearing nonlinearities

### 3.2.1 High-dimensional dynamic model of the system

In order to demonstrate the superior performance of the HB-AD method in capturing nonlinear responses of high-dimensional systems with complex nonlinearities, an aero-engine rotor-bearing-casing system[88] is considered here, as shown in **Fig. 4**. The rotor system primarily comprises four key components: namely the high-pressure rotor (HP rotor), low-pressure rotor (LP rotor), inner casing, and outer casing. By employing the finite element method, the high-dimensional nonlinear dynamic model of the system is established as follows:

$$\mathbf{M}\ddot{\mathbf{X}}+[\mathbf{C}+\mathbf{G}(\omega_1,\omega_2)]\dot{\mathbf{X}}+\mathbf{K}\mathbf{X}+\mathbf{F}_{\mathrm{B}}(\mathbf{X},t)=\mathbf{F}(\omega_1,\omega_2,t), \tag{30}$$

where **M**, **C**, **K** denote the system's mass matrix, damping matrix, and stiffness matrix, respectively. $\omega_1$ and $\omega_2$ represent the rotational speeds of the LP rotor and HP rotor, while $\mathbf{G}(\omega_1,\omega_2)$ denotes the gyroscopic matrix related to the rotational speeds of the LP rotor and HP rotor. $\mathbf{F}_{\mathrm{B}}(\mathbf{X},t)$ represents the nonlinear force of the inter-shaft bearing, calculated using the Hertz contact model, which incorporates three nonlinear factors: namely a fractional exponential function, clearance, and time-varying stiffness excitation. $\mathbf{F}(\omega_1,\omega_2,t)$ denotes the unbalance excitation forces originating from the LP rotor and HP rotor. For efficiency optimization in actual engineering applications, the LP rotor and HP rotor usually rotate at different speeds (denoted as $\omega_1$ and $\omega_2$ respectively). Consequently, $\mathbf{F}(\omega_1,\omega_2,t)$ encompasses two independent excitation frequencies corresponding to each rotor's rotational speed. More details about the dynamic model of the system can be found in our prior relevant work.

In summary, the rotor system's dynamics equations exhibit high degrees of freedom and complex nonlinear factors, while being subjected to dual-frequency excitation, rendering the solution of its dynamic response highly challenging. To address this, the HB-AD method is presented here.

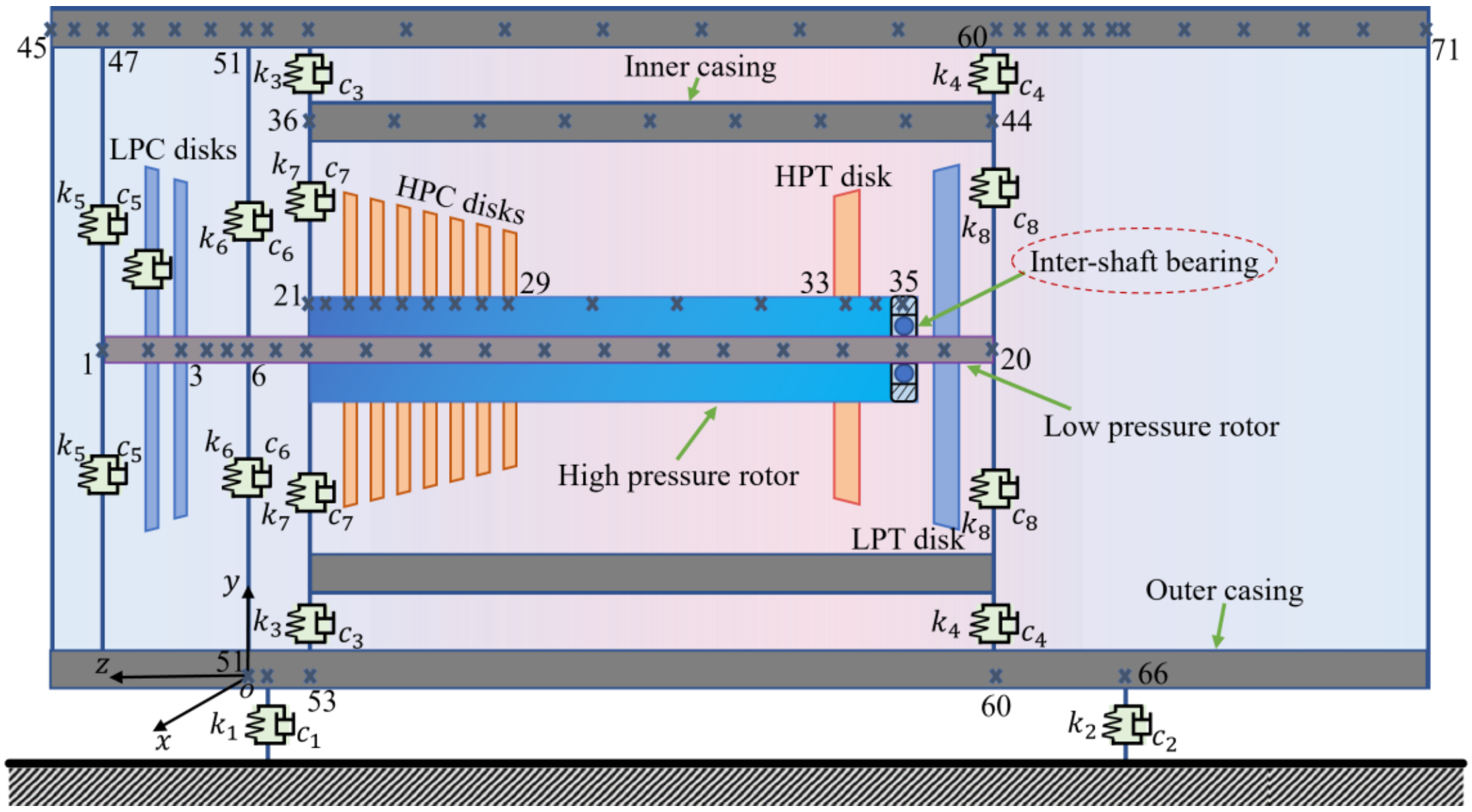

**Fig. 4** Schematic diagram of the aeroengine rotor system. The rotor system consists of four parts: namely the LP rotor, the HP rotor, the inner casing, and the outer casing. The finite element method is employed to establish the dynamic model, in which the LP rotor is divided with nodes 1-20, the HP rotor is divided with nodes 21-35, the inner casing is divided with nodes 36-44, and the outer casing is divided with 45-71. The model comprises a total of 71 nodes, each described by four degrees of freedom to characterize its motion. The system dynamics equations encompass a total of 284 degrees of freedom. The restoring force of the inter-shaft bearing is calculated using the Hertz contact model while incorporating three nonlinear factors: namely a fractional exponential function, clearance, and time-varying stiffness excitation.

### 3.2.2 Comparison of the HB-AD method, the HB-AFT method and the Newmark method

The system's dynamic equation (Eq. 30) was solved using the HB-AD method, HB-AFT method, and Newmark method to obtain the periodic responses near the first critical speed. **Fig. 5** illustrates the amplitude–frequency response of the LP rotor turbine disk (Node 19) with speed ratio ($\lambda = \omega_2 / \omega_1$) of 1.2. As shown in **Fig. 5**, the red solid line represents the stable periodic response obtained via the HB-AD method, while the red dashed line corresponds to its unstable counterpart. The blue and black circles indicate the stable and unstable responses computed using the HB-AFT method, respectively. The green stars denote the stable periodic response derived from the Newmark method. It is found that all three methods yield highly consistent results for the stable periodic response, thereby demonstrating the accuracy and reliability of the HB-AD method.

Furthermore, as shown in **Fig. 5**, the amplitude-frequency response of the rotor system exhibits two distinct resonance peaks, denoted as I and II. The first resonance peak is induced by the HP rotor passing through the rotor system’s critical speed, while the second corresponds to the LP rotor crossing the critical speed. Within both resonance regions, the system exhibits three periodic solutions: two stable and one unstable. The unstable periodic solutions obtained by the HB-AD and HB-AFT methods (coupled with the arc-length continuation technique) show excellent agreement, whereas the Newmark method fails to capture the system’s unstable responses. Compared to numerical integration approaches like the Newmark method, the HB-AD method retains the advantages of analytical techniques, enabling the computation of the complete set of periodic solutions—including unstable ones—thus providing a comprehensive view of the system’s dynamic behavior. This is essential for establishing a foundation for studying the underlying bifurcation mechanisms and stability characteristics of the system. In engineering practice, since unstable periodic solutions do not manifest in actual operation, the system tends to exhibit pronounced jump phenomena in the resonance regions, which are detrimental to its reliable and stable operation. Therefore, capturing the full spectrum of periodic responses is crucial for analyzing the underlying mechanisms of nonlinear phenomena such as vibration jumps and multi-solution behavior in the resonance zones.

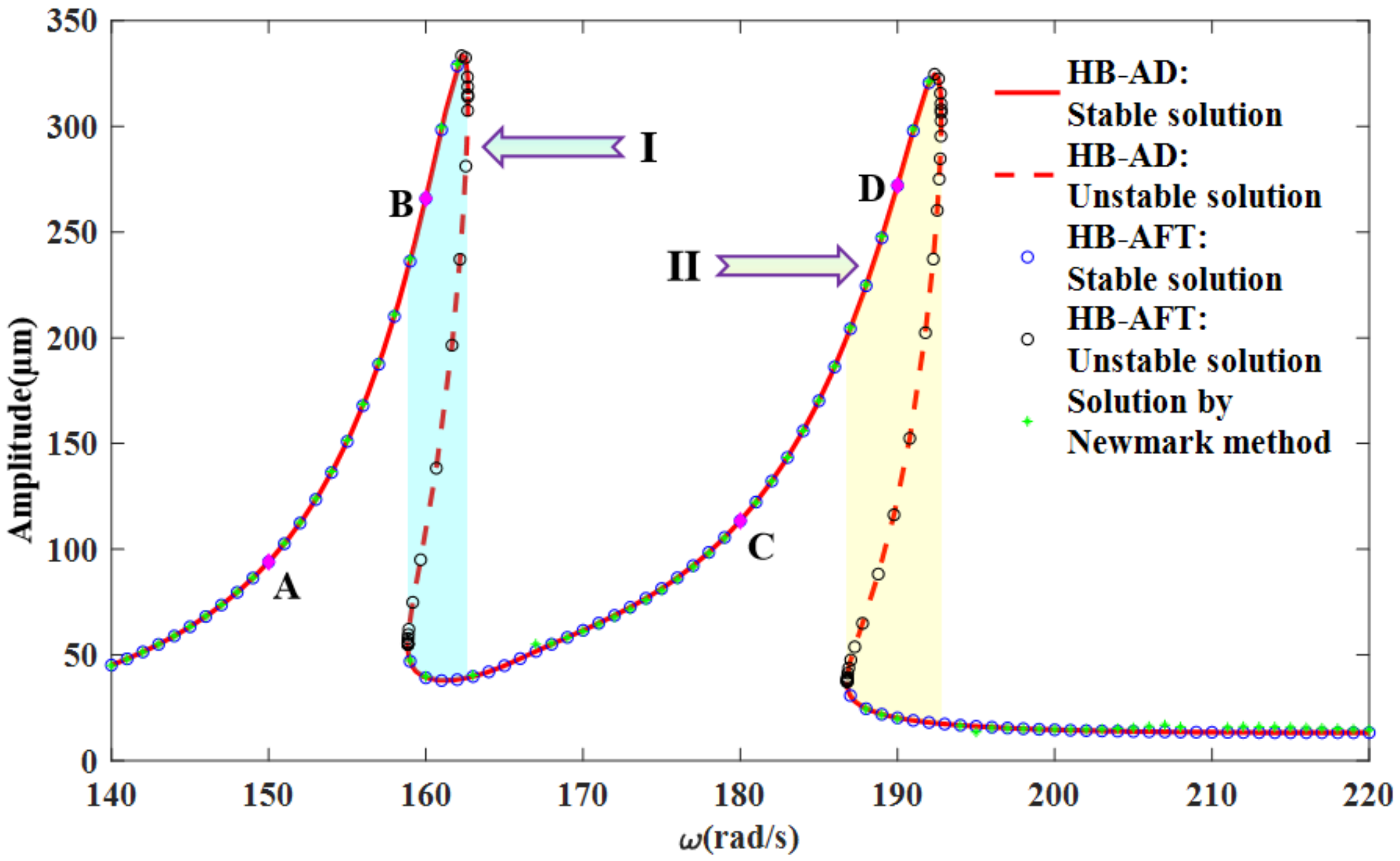

**Fig. 5** Amplitude-frequency response curve of the rotor system. There are two distinct resonance peaks, denoted as I and II. The system exhibits three periodic solutions: two stable and one unstable in the resonance regions. All three methods yield highly consistent results for the stable periodic response.

To further compare the results obtained by the three methods, **Fig. 6** and **Fig. 7** present the time history responses, rotor orbits, and frequency spectra of the rotor system in both resonant and non-resonant regions. Points A ($\omega = 150$ (rad/s)) and C ($\omega = 180$ (rad/s)) are selected from the non-resonant regions of the amplitude–frequency response curve, while Points B ($\omega = 160$ (rad/s)) and D ($\omega = 190$ (rad/s)) are taken from Resonance Regions I and II, respectively. These representative points sufficiently capture the full operational characteristics of the system across the amplitude–frequency domain. In the figures, the red solid lines represent the results computed by the HB-AD method, the blue circles indicate the results from the HB-AFT method, and the green circles correspond to those obtained by the Newmark method. As observed in **Fig. 6** and **Fig. 7**, the results from the three methods exhibit excellent agreement, thereby further validating the accuracy and effectiveness of the proposed approach.

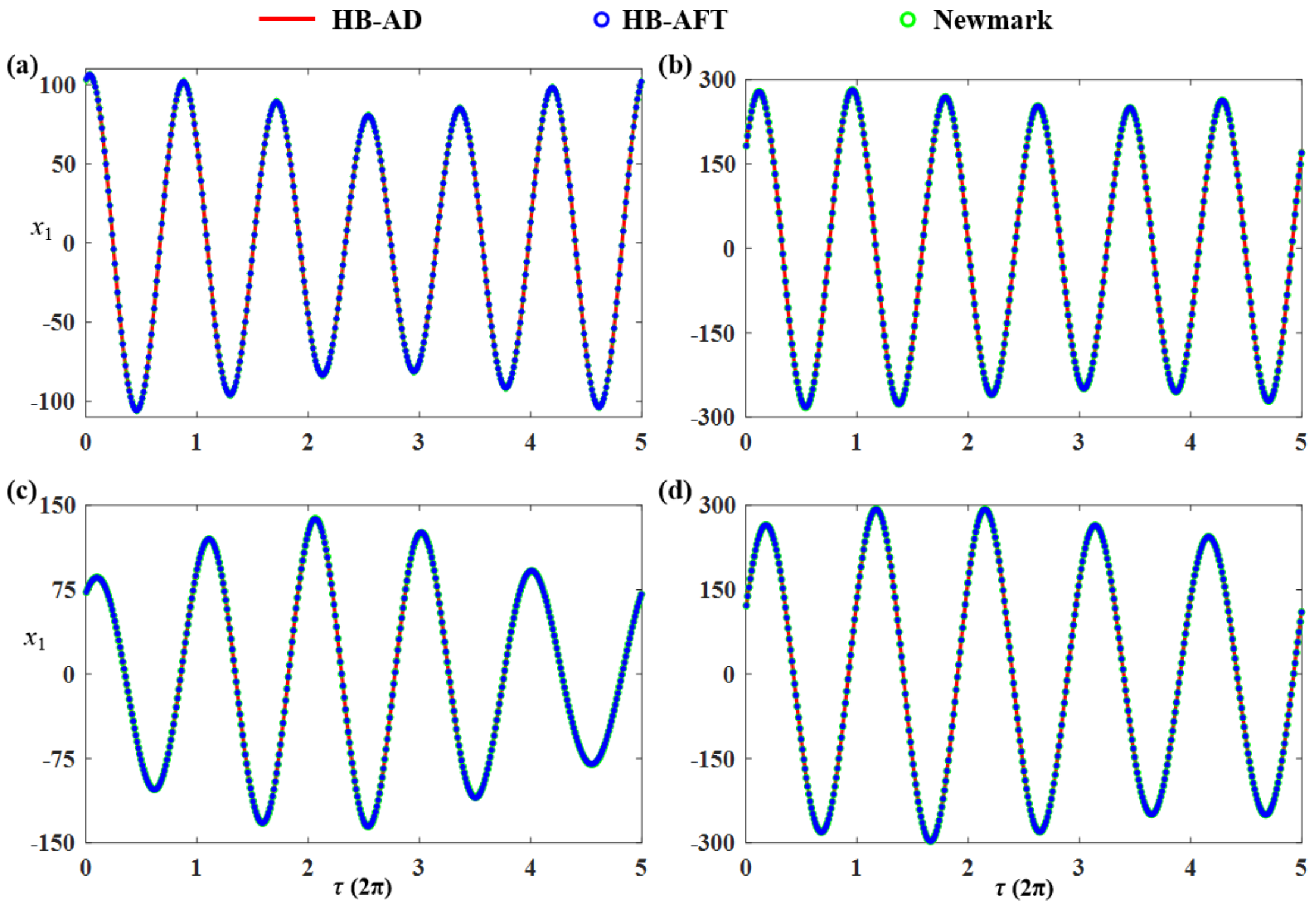

**Fig. 6** The time history diagram (node 19). (a) for point A, $\omega = 150$ (rad/s), in the non-resonant region; (b) for point B, at top of the resonance region I, $\omega = 160$ (rad/s); (c) for point C, $\omega = 180$ (rad/s), in the non-resonant region; (d) for point D, at top of the resonance region II, $\omega = 190$ (rad/s)

To further compare the computational performance of the HB-AD method, the HB-AFT method, and the Newmark method, **Table. 2** presents the computational times required by each method at points A, B, C, and D. For each successive frequency point, the initial guess solution was taken as the converged response from the previous frequency point. The frequency step increment was uniformly set to 1 rad/s across all calculations. All methods employed the identical convergence criterion for their Newton iterations, specifically an error tolerance of $1 \times 10^{-9}$. As summarized in Table 1, the mean computational time required by the Newmark method to compute the periodic responses across the four points (A, B, C, D) was 820.03 seconds. In contrast, the HB-AFT method exhibited a mean computational time of 99.04 seconds, while the HB-AD method achieved a significantly reduced mean time of 5.70 seconds. In other words, the HB-AD method demonstrates an average computational efficiency approximately 17 times higher than the HB-AFT method and about 144 times higher than the Newmark method. Overall, the HB-AD method exhibits a substantial efficiency advantage over both the HB-AFT and Newmark methods.This enhanced computational efficiency will be critical for facilitating large-scale parametric studies in engineering practice, enabling the systematic analysis of parameter influence laws necessary for the effective structural parameter optimization design of complex rotor systems.

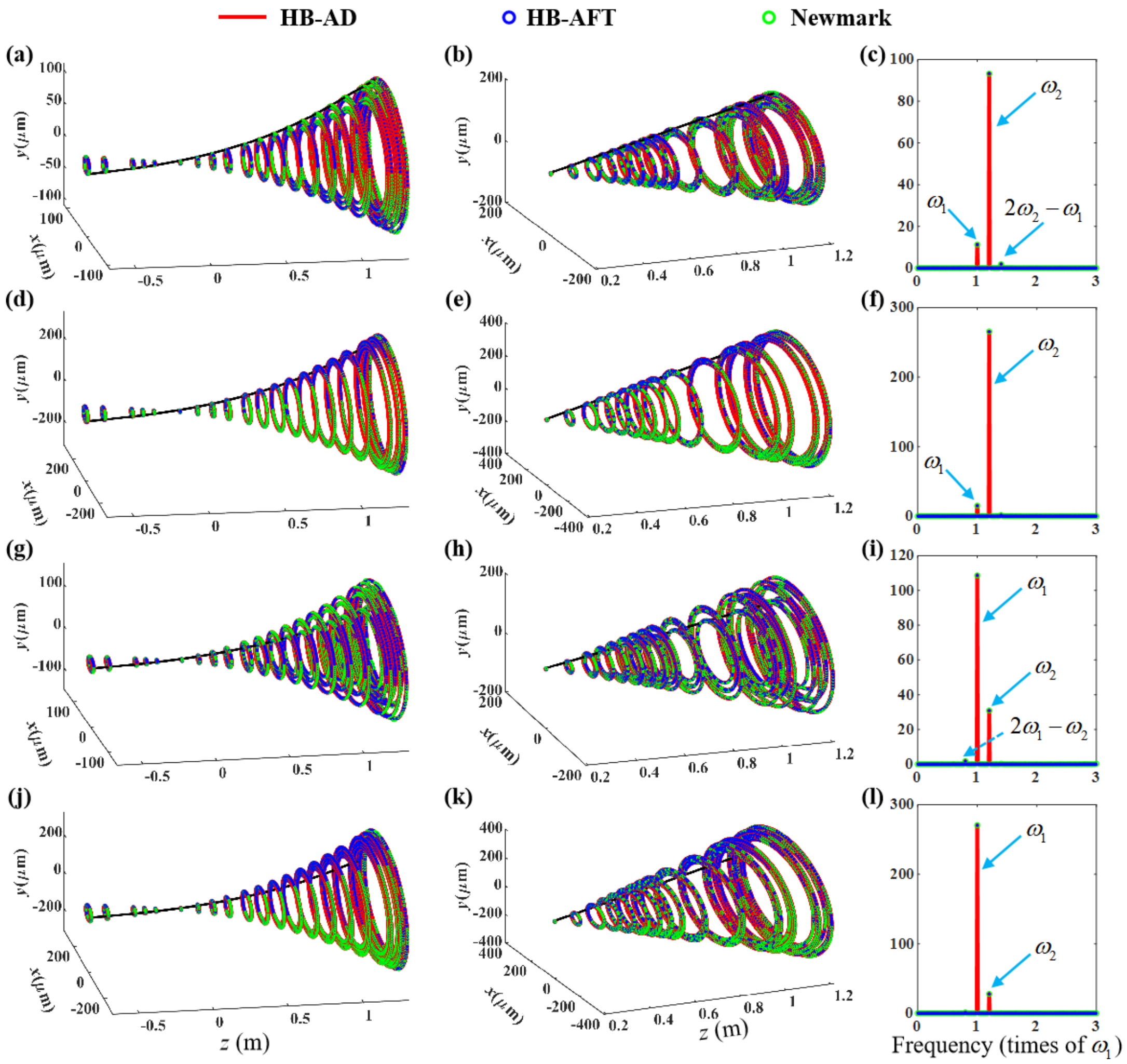

**Fig. 7** The rotors' orbits and frequency spectrum (node 19). (a) rotors' orbits of the LP rotor at point A, (b) rotors' orbits of the HP rotor at point A, (c) frequency spectrum at point A; (d) rotors' orbits of the LP rotor at point B, (e) rotors' orbits of the HP rotor at point B, (f) frequency spectrum at point B; (g) rotors' orbits of the LP rotor at point C, (h) rotors' orbits of the HP rotor at point C, (i) frequency spectrum at point C; (j) rotors' orbits of the LP rotor at point D, (k) rotors' orbits of the HP rotor at point D, (l) frequency spectrum at point D.

**Table. 2** Comparison of computation time between the HB-AD method, the HB-AFT method and the Newmark method.

| Points of comparison | Computation time (s) | | |
|---|---|---|---|
| | Newmark method | HB-AFT method | HB-AD method |
| A ( $\omega = 150$ (rad/s) ) | 817.66 | 89.56 | 4.68 |
| B ( $\omega = 160$ (rad/s) ) | 821.95 | 108.11 | 6.30 |
| C ( $\omega = 180$ (rad/s) ) | 817.75 | 90.05 | 5.47 |
| D ( $\omega = 190$ (rad/s) ) | 822.76 | 108.42 | 6.34 |

# 4 Discussion

Notably, the HB-AD method shares identical mathematical principles and theoretical frameworks with traditional semi-analytical methods based on Harmonic balance and Newton iteration (e.g., HB-AFT and IHB methods). Fundamentally, all these approaches employ Harmonic balance to transform nonlinear system differential equations into algebraic equations, followed by Newton iteration to solve these equations and obtain periodic responses. The computation of the Jacobian matrix is the critical determinant of performance. Although significant advances have been made in optimizing Jacobian calculations—such as leveraging Fast Fourier Transform, dimension reduction, and tensor contraction—these improvements remain confined to symbolic or numerical differentiation paradigms. This limitation necessitates extensive manual derivation of formulas, which is not only labor-intensive and error-prone but also prohibitively challenging for complex systems. For systems with intricate nonlinearities (e.g., Rotor system with SFD, shown in section 3.1): symbolic differentiation often induces expression swell, making analytical derivation of the Jacobian matrix practically intractable. For high-dimensional systems (e.g., Aeroengine rotor system with bearing nonlinearities, shown in section 3.2): even if Jacobian expressions are manually derived, precise alignment of matrix elements with system's degrees of freedom (DOFs) and Fourier-series orders of periodic solutions demands meticulous programming. This process is cumbersome, time-consuming, and susceptible to implementation errors. Numerical differentiation alternatives introduce truncation/round-off errors, precluding exact Jacobian computation. Near bifurcation points—where nonlinear effects dominate (e.g., in the resonance region of aeroengine rotor system with bearing nonlinearities, shown in section 3.2)—these errors impede convergence and obscure the comprehensive characterization of periodic responses.

The HB-AD method inherits the advantages of semi-analytical approaches based on harmonic balance and Newton iteration while integrating automatic differentiation to overcome the limitations of cumbersome formula derivation and limited generalizability inherent in such methods. Its core innovation lies in a unified solution framework that leverages AD to achieve generic, efficient, and precise computation of the Jacobian matrix, eliminating all manual derivation processes. This effectively resolves the intrinsic limitations of both symbolic differentiation (e.g., expression swell in complex systems) and numerical differentiation (e.g., truncation/round-off errors), thereby significantly enhancing the overall performance of the solution methodology and establishing a robust foundation for universal implementation. Furthermore, the HB-AD method exhibits substantial potential for subsequent adoption and development into a general-purpose nonlinear system analysis software. Within this framework, users can efficiently compute periodic responses by inputting only key system elements—such as mass, damping, stiffness, external excitation, and nonlinear terms—without requiring specialized knowledge or manual formula manipulation. This enables truly "out-of-the-box" usability, democratizing access to high-fidelity solutions for complex dynamical systems.

While the HB-AD method demonstrates significant computational efficiency advantages over the Newmark method and traditional HB-AFT method, as shown in section 3.2.2, it is noteworthy that employing automatic differentiation inevitably incurs computational overhead. This arises because the Jacobian matrix must be recomputed during each Newton-Raphson iteration cycle. In contrast, HB-AFT and IHB method, when integrated with Fast Fourier Transform and dimension reduction, achieve substantially enhanced efficiency in Jacobian matrix computation, thereby boosting overall performance. The current authors' prior work[88] employed an enhanced HB-AFT approach (termed GE-HB) to analyze the aeroengine rotor system with bearing nonlinearities (as shown in section 3.2), revealing that GE-HB outperforms HB-AD by approximately an order of magnitude in computational speed for single-parameter cases. Nevertheless, this efficiency trade-off is acceptable for two critical reasons. First, both GE-HB and HB-AD exhibit high computational efficiency; marginally slower convergence in HB-AD does not materially impede practical workflows. More importantly, HB-AD leverages parallel computing capabilities and CUDA acceleration via advanced deep learning frameworks (e.g., PyTorch, JAX). Consequently, while HB-AD may lag behind GE-HB in single-parameter analyses, it excels in large-scale parametric studies typical of engineering applications. The parallelizable architecture of HB-AD enables simultaneous computation across multiple parameter sets, drastically reducing total solution time for high-dimensional problems. In summary, the HB-AD method achieves a remarkable equilibrium among generality, usability, and efficiency, with the first two attributes being particularly salient. Its out-of-the-box usability—requiring minimal user-derived equations—offers substantial practical benefits to researchers and engineers tackling complex nonlinear dynamical systems.

# 5 Conclusions

In this paper, the HB-AD method is proposed by integrating AD with harmonic balance-Newton iteration framework. This approach automates the computation of Jacobian matrices required for Newton iterations by AD, enabling universal and efficient calculation of periodic responses in high-dimensional complex nonlinear systems. Through two representative case studies, the superior performance of HB-AD is demonstrated: specifically, in out-of-the-box usability, effective handling of complex nonlinear factors effectively, and high computational efficiency.

The HB-AD method adopts AD to compute Jacobian matrices without the need for manual derivation, resolving limitations of symbolic differentiation (expression explosion) and numerical differentiation (truncation/rounding errors). This enables out-of-the-box usability and enhances generality for complex nonlinear systems. Furthermore, leveraging advanced deep learning frameworks and native CUDA support, the HB-AD method designs an optimized program architecture that directly invokes the reverse-mode AD function for efficient Jacobian matrix computation. By integrating arc-length continuation and AD, HB-AD method autonomously obtain the complete set of periodic responses—including both stable and unstable solution branches—directly from user-supplied governing dynamic equations, eliminating the need for manual derivation of analytical formulations.These implementations enable out-of-the-box deployment and significantly enhances computational efficiency.

The results of two numerical case studies demonstrate the superior performance of the HB-AD method for dynamic characteristic analysis of complex nonlinear systems. The first case study involves a rotor system with SFD where oil film forces contain complex functional forms, and here the HB-AD method achieves efficient computation of its periodic responses, confirming excellent applicability to such systems. The second case study pertains to an aero-engine rotor-bearing-casing system with 284 DOFs, incorporating complex bearing nonlinearities and subjected to dual-frequency excitation, computational results show that the HB-AD method captures all periodic solutions (including unstable periodic responses) while maintaining high efficiency—specifically, it is 17 times faster than the conventional HB-AFT method and 144 times faster than the Newmark method.

This methodology advancement establishes a robust methodology framework for rapid acquisition of dynamic responses in high-dimensional complex nonlinear systems for practical engineering applications.

## Conflict of Interest

The authors declare that they have no conflict of interest.

## Code and data availability statements

The proposed methodology of this study is openly available in GitHub at https://github.com/shuizidesu/hb-ad.

## Acknowledgments

The authors are very grateful for the financial supports from National Natural Science Foundation of China (Grant Nos. 12422213, 12372008, U244120491), and National Key R&D Program of China (Grant No. 2023YFE0125900).

## References

[1] Dong Y, Hu H, Wang L. A comprehensive study on the coupled multi-mode vibrations of cylindrical shells. Mechanical Systems and Signal Processing 2022;169:108730. https://doi.org/10.1016/j.ymssp.2021.108730.

[2] Gu J, Chen X, Jung J-H. Explicit radial basis function Runge–Kutta methods. Numer Algor 2025. https://doi.org/10.1007/s11075-025-02121-w.

[3] Nair P, Vignesh D. A study on generalized balanced split drift stochastic Runge- Kutta methods for stochastic differential equations. Phys Scr 2024;99:115249. https://doi.org/10.1088/1402-4896/ad7f0a.

[4] Hou Y, Cao S, Kang Y, Li G. Dynamics analysis of bending–torsional coupling characteristic frequencies in dual-rotor systems. AIAA Journal 2022;60:6020–35.

https://doi.org/10.2514/1.J061848.
[5] Zhou J, Luo Z, Li L, Ma T, Li H. Numerical and experimental analysis of the influence of elastic supports on bearing-rotor systems. Mech Syst Signal Process 2025;224:112235. https://doi.org/10.1016/j.ymssp.2024.112235.
[6] Kim VA, Parovik RI. Application of the Explicit Euler Method for Numerical Analysis of a Nonlinear Fractional Oscillation Equation. Fractal Fract 2022;6:274. https://doi.org/10.3390/fractalfract6050274.
[7] Koellermeier J, Samaey G. Projective integration methods in the Runge–Kutta framework and the extension to adaptivity in time. Journal of Computational and Applied Mathematics 2025;454:116147. https://doi.org/10.1016/j.cam.2024.116147.
[8] Zhang H, Yan J, Qian X, Song S. Temporal high-order, unconditionally maximum-principle-preserving integrating factor multi-step methods for Allen-Cahn-type parabolic equations. Applied Numerical Mathematics 2023;186:19–40. https://doi.org/10.1016/j.apnum.2022.12.020.
[9] Hu Q, Deng S, Teng H. A 5-DOF Model for Aeroengine Spindle Dual-rotor System Analysis. Chinese Journal of Aeronautics 2011;24:224–34. https://doi.org/10.1016/s1000-9361(11)60027-7.
[10] Liu J, Wang C, Luo Z. Research nonlinear vibrations of a dual-rotor system with nonlinear restoring forces. J Braz Soc Mech Sci Eng 2020;42:461. https://doi.org/10/gn79z5.
[11] Wang F, Luo G-H, Yan S, Cui H-T. A Comparison Study on Co- and Counterrotating Dual-Rotor System with Squeeze Film Dampers and Intermediate Bearing. Shock and Vibration 2017;2017:1–25. https://doi.org/10.1155/2017/5493763.
[12] Ma X, Ma H, Qin H, Guo X, Zhao C, Yu M. Nonlinear vibration response characteristics of a dual-rotor-bearing system with squeeze film damper. Chinese Journal of Aeronautics 2021;34:128–47. https://doi.org/10.1016/j.cja.2021.01.013.
[13] Zhang ZJ, Dong YY, Han YW. On the predictive modeling of nonlinear frequency behaviors of an archetypal rub-impact rotor. International Journal of Mechanical Sciences 2019;161–162:105083. https://doi.org/10.1016/j.ijmecsci.2019.105083.
[14] Yang Y, Cao D, Yu T, Wang D, Li C. Prediction of dynamic characteristics of a dual-rotor system with fixed point rubbing—Theoretical analysis and experimental study. International Journal of Mechanical Sciences 2016;115–116:253–61. https://doi.org/10.1016/j.ijmecsci.2016.07.002.
[15] Wang D, Cao H, Yang Y, Du M. Dynamic modeling and vibration analysis of cracked rotor-bearing system based on rigid body element method. Mechanical Systems and Signal Processing 2023;191:110152. https://doi.org/10.1016/j.ymssp.2023.110152.
[16] Cao L, Xue C, Si H. Research on the dynamic characteristics of rotor-bearing-seal system with breathing cracks under steam flow excited force. International Journal of Non-Linear Mechanics 2023;157:104546. https://doi.org/10.1016/j.ijnonlinmec.2023.104546.
[17] Xu H, Yang Y, Ma H, Luo Z, Li X, Han Q, et al. Vibration characteristics of bearing-rotor systems with inner ring dynamic misalignment. International Journal of Mechanical Sciences 2022;230:107536. https://doi.org/10.1016/j.ijmecsci.2022.107536.
[18] Wang N, Jiang D. Vibration response characteristics of a dual-rotor with unbalance-misalignment coupling faults: Theoretical analysis and experimental study. Mechanism and Machine Theory 2018;125:207–19. https://doi.org/10/gdnw9s.
[19] Hou L, Yi H, Jin Y, Gui M, Sui L, Zhang J, et al. Inter-shaft Bearing Fault Diagnosis Based on Aero-engine System: A Benchmarking Dataset Study. JDMD 2023;2:228–42.

https://doi.org/10.37965/jdmd.2023.314.
[20] Gao T, Cao S. Paroxysmal impulse vibration phenomena and mechanism of a dual–rotor system with an outer raceway defect of the inter-shaft bearing. Mechanical Systems and Signal Processing 2021;157:107730. https://doi.org/10/gh7h3d.
[21] Zeng Y, Huang P, He Y. A time filter method for solving the double-diffusive natural convection model. Computers & Fluids 2022;235:105265. https://doi.org/10.1016/j.compfluid.2021.105265.
[22] OGUNFEYITIMI SE, IKHILE MNO. Implicit-explicit second derivative lmm for stiff ordinary differential equations. Journal of the Korean Society for Industrial and Applied Mathematics 2021;25:224–61. https://doi.org/10.12941/JKSIAM.2021.25.224.
[23] Sotiropoulos F, Constantinescu G. Pressure-Based Residual Smoothing Operators for Multistage Pseudocompressibility Algorithms. Journal of Computational Physics 1997;133:129–45. https://doi.org/10.1006/jcph.1997.5662.
[24] Ma X, Li Z, Xiang J, Chen C, Huang F. Vibration characteristics of rotor system with coupling misalignment and disc-shaft nonlinear contact. Mechanical Systems and Signal Processing 2025;223:111839. https://doi.org/10.1016/j.ymssp.2024.111839.
[25] Jiang Y, Jin Y, Hou L, Chen Y, Cong A. A modified Newmark/Newton-Raphson method with automatic differentiation for general nonlinear dynamics analysis 2025. https://doi.org/10.48550/arXiv.2506.13226.
[26] Zhou J, Luo Z, Li L, Tang R, Ma T, Yang D. A dynamic model of a three-point contact ball bearing-rotor system: numerical and experimental verification. Nonlinear Dyn 2025;113:14471–96. https://doi.org/10.1007/s11071-025-10880-3.
[27] Liu K, Shi X, Wang D, Feng Y, Jian Y, Li W. A method for the dynamic characteristic analysis of a rotor-rolling bearing system influenced by elastohydrodynamic lubrication. Journal of Sound and Vibration 2025;608:119075. https://doi.org/10.1016/j.jsv.2025.119075.
[28] Zhang B, Chen X, Xiang F, Ren G, Gan X. Dynamic Characteristics and Periodic Stability Analysis of Rotor System with Squeeze Film Damper Under Base Motions. Applied Sciences 2025;15:1186. https://doi.org/10.3390/app15031186.
[29] Wang J, Zhang X, Liu Y, Qin Z, Ma L, Chu F. Rotor vibration control via integral magnetorheological damper. International Journal of Mechanical Sciences 2023;252:108362. https://doi.org/10.1016/j.ijmecsci.2023.108362.
[30] Hou S, Lin R, Hou L, Chen Y. Dynamic characteristics of a dual-rotor system with parallel non-concentricity caused by inter-shaft bearing positioning deviation. Mechanism and Machine Theory 2023;184:105262. https://doi.org/10.1016/j.mechmachtheory.2023.105262.
[31] Wu Z, Hao L, Zhao W, Ma Y, Bai S, Zhao Q. Modeling and vibration analysis of an aero-engine dual-rotor-support-casing system with inter-shaft rub-impact. International Journal of Non-Linear Mechanics 2024;165:104757. https://doi.org/10.1016/j.ijnonlinmec.2024.104757.
[32] Miao X, He J, Zhang D, Jiang D, Li J, Ai X, et al. Nonlinear response analysis of variable speed rotor system under maneuvering flight. J Mech Sci Technol 2023;37:4957–71. https://doi.org/10.1007/s12206-023-0903-x.
[33] Pan W, Hao J, Li H, Wang J, Bao J, Zeng X, et al. Dynamic modeling and response analysis of misaligned rotor system with squeeze film dampers under maneuver loads. Nonlinear Dyn 2025;113:189–233. https://doi.org/10.1007/s11071-024-09862-8.
[34] Yang CH, Zhu SM, Chen SH. A modified elliptic Lindstedt–Poincaré method for certain strongly non-linear oscillators. Journal of Sound and Vibration 2004;273:921–32.

https://doi.org/10.1016/S0022-460X(03)00565-0.
[35] Awrejcewicz J, Sypniewska-Kamińska G, Mazur O. Analysing regular nonlinear vibrations of nano/micro plates based on the nonlocal theory and combination of reduced order modelling and multiple scale method. Mechanical Systems and Signal Processing 2022;163:108132. https://doi.org/10.1016/j.ymssp.2021.108132.
[36] Wang Q, Yan Z, Dai H. An efficient multiple harmonic balance method for computing quasi-periodic responses of nonlinear systems. Journal of Sound and Vibration 2023;554:117700. https://doi.org/10.1016/j.jsv.2023.117700.
[37] Chang Z, Hou L, Chen Y. Investigation on the 1:2 internal resonance of an FGM blade. Nonlinear Dyn 2022;107:1937–64. https://doi.org/10.1007/s11071-021-07070-2.
[38] Ma N-J, Wang R-H, Han Q. Primary parametric resonance–primary resonance response of stiffened plates with moving boundary conditions. Nonlinear Dyn 2015;79:2207–23. https://doi.org/10.1007/s11071-014-1806-2.
[39] Chen Y, Hou L, Chen G, Song H, Lin R, Jin Y, et al. Nonlinear dynamics analysis of a dual-rotor-bearing-casing system based on a modified HB-AFT method. Mech Syst Signal Process 2023;185:109805. https://doi.org/10.1016/j.ymssp.2022.109805.
[40] Hou L, Chen Y, Chen Y. Combination resonances of a dual-rotor system with inter-shaft bearing. Nonlinear Dyn 2023;111:5197–219. https://doi.org/10.1007/s11071-022-08133-8.
[41] Chen Y, Hou L, Lin R, Wang Y, Saeed NA, Chen Y. Combination resonances of a dual-rotor-bearing-casing system. Nonlinear Dyn 2024;112:4063–83. https://doi.org/10.1007/s11071-024-09282-8.
[42] Xu XHL. Internal Resonance Analysis for Electromechanical Integrated Toroidal Drive. Journal of Computational and Nonlinear Dynamics 2010;5. https://doi.org/10.1115/1.4001821.
[43] Hong D, Hill TL, Neild SA. Existence and location of internal resonance of two-mode nonlinear conservative oscillators. Proc R Soc A 2022;478. https://doi.org/10.1098/rspa.2021.0659.
[44] Yang R, Jin Y, Hou L, Chen Y. Advantages of pulse force model over geometrical boundary model in a rigid rotor–ball bearing system. International Journal of Non-Linear Mechanics 2018;102:159–69. https://doi.org/10.1016/j.ijnonlinmec.2018.03.011.
[45] Meng Q, Hou L, Lin R, Chen Y, Cui G, Shi W, et al. Accurate nonlinear dynamic characteristics analysis of quasi-zero-stiffness vibration isolator via a modified incremental harmonic balance method. Nonlinear Dyn 2024;112:125–50. https://doi.org/10.1007/s11071-023-09036-y.
[46] Lu K, Chen Y, Hou L. Bifurcation characteristics analysis of a class of nonlinear dynamical systems based on singularity theory. Appl Math Mech-Engl Ed 2017;38:1233–46. https://doi.org/10.1007/s10483-017-2234-8.
[47] Ma S-L, Huang T, Yan Y, Zhang X-M, Ding H, Wiercigroch M. Bifurcation analysis of thin-walled structures trimming process with state-dependent time delay. International Journal of Mechanical Sciences 2024;271:109159. https://doi.org/10.1016/j.ijmecsci.2024.109159.
[48] Chen SH, Cheung YK. A modified lindstedt–poincare method for a strongly non-linear two degree-of-freedom system. Journal of Sound and Vibration 1996;193:751–62. https://doi.org/10.1006/jsvi.1996.0313.
[49] Chen SH, Cheung YK. An Elliptic Lindstedt–Poincare´ Method for Certain Strongly Non-Linear Oscillators. Nonlinear Dynamics 1997;12:199–213.
[50] Tam KK. On the asymptotic solution of the Orr–Sommerfeld equation by the method of

multiple-scales. J Fluid Mech 1968;34:145–58. https://doi.org/10.1017/S0022112068001801.

[51] Chaturvedi S, Gardiner CW. Resonance fluorescence of a two-level atom driven by a fluctuating laser field: an application of a multiple time scale method. J Phys B: At Mol Phys 1981;14:1119–38. https://doi.org/10.1088/0022-3700/14/7/009.

[52] Raze G, Volvert M, Kerschen G. Tracking amplitude extrema of nonlinear frequency responses using the harmonic balance method. Numerical Meth Engineering 2023:e7376. https://doi.org/10.1002/nme.7376.

[53] Wu J, Hong L, Jiang J. A robust and efficient stability analysis of periodic solutions based on harmonic balance method and Floquet-Hill formulation. Mechanical Systems and Signal Processing 2022;173:109057. https://doi.org/10.1016/j.ymssp.2022.109057.

[54] Xu M, Cai B, Li C, Zhang H, Liu Z, He D, et al. Dynamic characteristics and reliability analysis of ball screw feed system on a lathe. Mechanism and Machine Theory 2020;150:103890. https://doi.org/10.1016/j.mechmachtheory.2020.103890.

[55] Yang X-D, Zhang W. Nonlinear dynamics of axially moving beam with coupled longitudinal–transversal vibrations. Nonlinear Dyn 2014;78:2547–56. https://doi.org/10.1007/s11071-014-1609-5.

[56] Lou J, Fang X, Du J, Wu H. Propagation of fundamental and third harmonics along a nonlinear seismic metasurface. International Journal of Mechanical Sciences 2022;221:107189. https://doi.org/10.1016/j.ijmecsci.2022.107189.

[57] Zheng C, Gao J, Liu J, Xue X. Dynamic Performances of a Double-Layer Vibration Isolation System: Nonlinear Modeling and Experimental Validation. Int Journal of Mech Sys Dyn 2025;5:113–28. https://doi.org/10.1002/msd2.12138.

[58] Liu C-S, Kuo C-L, Chang C-W. Linearized Harmonic Balance Method for Seeking the Periodic Vibrations of Second- and Third-Order Nonlinear Oscillators. Mathematics 2025;13:162. https://doi.org/10.3390/math13010162.

[59] Dong Y, Wang Z, Mao X, Amabili M. On the combined iteration scheme of harmonic balance-pseudo arclength-extrapolation method applied to the nonlinear systems with nonlinear damping and time delay. Nonlinear Dyn 2025;113:20729–46. https://doi.org/10.1007/s11071-025-11284-z.

[60] Leung AYT, Chui SK. Non-linear vibration of coupled duffing oscillators by an improved incremental harmonic balance method. Journal of Sound and Vibration 1995;181:619–33. https://doi.org/10.1006/jsvi.1995.0162.

[61] Ma Q, Kahraman A. Period-one motions of a mechanical oscillator with periodically time-varying, piecewise-nonlinear stiffness. J Sound Vib 2005;284:893–914. https://doi.org/10.1016/j.jsv.2004.07.026.

[62] Chen S, Wang Y, Wu Q, Zhang X, Cao D, Wang B. Autonomous vibration control of beams utilizing intelligent excitation adaptability. International Journal of Mechanical Sciences 2025;293:110194. https://doi.org/10.1016/j.ijmecsci.2025.110194.

[63] Wu Q, Wang Y, Cao D. Nonlinear dynamic analysis of high aspect ratio wings via IHB method. Nonlinear Dynamics 2025;113:16225–44.

[64] Zhang Z, Chen Y, Li Z. Influencing factors of the dynamic hysteresis in varying compliance vibrations of a ball bearing. Sci China Technol Sci 2015;58:775–82. https://doi.org/10.1007/s11431-015-5808-1.

[65] Zhang Z, Chen Y, Cao Q. Bifurcations and hysteresis of varying compliance vibrations in the primary parametric resonance for a ball bearing. Journal of Sound and Vibration 2015;350:171–84. https://doi.org/10.1016/j.jsv.2015.04.003.

[66] Li H, Chen Y, Hou L, Zhang Z. Periodic response analysis of a misaligned rotor system by harmonic balance method with alternating frequency/time domain technique. Sci China Technol Sci 2016;59:1717–29. https://doi.org/10/f9btd9.

[67] Sun W, Jiang Z, Xu X, Han Q, Chu F. Harmonic balance analysis of output characteristics of free-standing mode triboelectric nanogenerators. International Journal of Mechanical Sciences 2021;207:106668. https://doi.org/10.1016/j.ijmecsci.2021.106668.

[68] Zheng Z, Lu Z, Liu G, Chen Y. Twice Harmonic Balance Method for Stability and Bifurcation Analysis of Quasi-Periodic Responses. Journal of Computational and Nonlinear Dynamics 2022;17:121006. https://doi.org/10.1115/1.4055923.

[69] Pei L, Chong ASE, Pavlovskaia E, Wiercigroch M. Computation of periodic orbits for piecewise linear oscillator by Harmonic Balance Methods. Communications in Nonlinear Science and Numerical Simulation 2022;108:106220. https://doi.org/10.1016/j.cnsns.2021.106220.

[70] Zhang K, Lu K, Chai S, Cheng H, Fu C, Guo D. Dynamic modeling and parameter sensitivity analysis of AUV by using the POD method and the HB-AFT method. Ocean Engineering 2024;293:116693. https://doi.org/10.1016/j.oceaneng.2024.116693.

[71] Lin R, Hou L, Dun S, Cai Y, Sun C, Chen Y. Synchronous impact phenomenon of a high-dimension complex nonlinear dual-rotor system subjected to multi-frequency excitations. Sci China Technol Sci 2023;66:1757–68. https://doi.org/10.1007/s11431-022-2215-0.

[72] Ma X, Zhang Z, Hua H. Uncertainty quantization and reliability analysis for rotor/stator rub-impact using advanced Kriging surrogate model. Journal of Sound and Vibration 2022;525:116800. https://doi.org/10.1016/j.jsv.2022.116800.

[73] Ouyang X. Nonlinear dynamics of a dual-rotor-bearing system with active elastic support dry friction dampers n.d.

[74] Zhang L, Yuan Q, Ma X, Hou X, Li Z, Zhang J, et al. Research on rub-impact rotor vibration of hydraulic generating set based on HB-AFT method. Nonlinear Dyn n.d. https://doi.org/10.1007/s11071-022-08199-4.

[75] Wang D, Song L, Zhu R, Cao P. Nonlinear dynamics and stability analysis of dry friction damper for supercritical transmission shaft. Nonlinear Dyn 2022;110:3135–49. https://doi.org/10.1007/s11071-022-07795-8.

[76] Tang Z, Zhu H, Luo H, Li T. Performance of a force-restricted viscous mass damper incorporated into base-isolated liquid storage tanks. Structures 2024;61:106002. https://doi.org/10.1016/j.istruc.2024.106002.

[77] Wang Q, Zhang Z, Ying Y, Pang Z. Analysis of the Dynamic Stiffness, Hysteresis Resonances and Complex Responses for Nonlinear Spring Systems in Power-Form Order. Applied Sciences 2021;11:7722. https://doi.org/10.3390/app11167722.

[78] Coudeyras N, Sinou J-J, Nacivet S. A new treatment for predicting the self-excited vibrations of nonlinear systems with frictional interfaces: The Constrained Harmonic Balance Method, with application to disc brake squeal. Journal of Sound and Vibration 2009;319:1175–99. https://doi.org/10.1016/j.jsv.2008.06.050.

[79] Wu J, Hong L, Jiang J. A comparative study on multi- and variable-coefficient harmonic balance methods for quasi-periodic solutions. Mechanical Systems and Signal Processing 2023;187:109929. https://doi.org/10.1016/j.ymssp.2022.109929.

[80] Chopra I, Dugundji J. Non-linear dynamic response of a wind turbine blade. Journal of Sound and Vibration 1979;63:265–86. https://doi.org/10.1016/0022-460X(79)90883-6.

[81] Lau SL, Cheung YK. Amplitude Incremental Variational Principle for Nonlinear Vibration of

Elastic Systems. Journal of Applied Mechanics 1981;48:959–64. https://doi.org/10.1115/1.3157762.

[82] Kim YB, Noah ST. Stability and bifurcation analysis of oscillators with piecewise-linear characteristics: a general approach. Journal of Applied Mechanics 1991;58:545–53. https://doi.org/10/cngcz6.

[83] Carrera E. A study on arc-length-type methods and their operation failures illustrated by a simple model. Computers & Structures 1994;50:217–29. https://doi.org/10.1016/0045-7949(94)90297-6.

[84] Ritto-Correa M, Camotim D. On the arc-length and other quadratic control methods: Established, less known and new implementation procedures. Computers & Structures 2008;86:1353–68. https://doi.org/10.1016/j.compstruc.2007.08.003.

[85] Guskov M, Thouverez F. Harmonic balance-based approach for quasi-periodic motions and stability analysis. J Vib Acoust-Trans ASME 2012;134:031003. https://doi.org/10.1115/1.4005823.

[86] Ge T, Leung AYT. A Toeplitz Jacobian Matrix/Fast Fourier Transformation Method for Steady-State Analysis of Discontinuous Oscillators. Shock and Vibration 1995;2:205–18. https://doi.org/10.1155/1995/973431.

[87] Guskov M, Sinou J-J, Thouverez F. Multi-dimensional harmonic balance applied to rotor dynamics. Mech Res Commun 2008;35:537–45. https://doi.org/10.1016/j.mechrescom.2008.05.002.

[88] Chen Y, Hou L, Lin R, Toh W, Ng TY, Chen Y. A general and efficient harmonic balance method for nonlinear dynamic simulation. International Journal of Mechanical Sciences 2024;276:109388. https://doi.org/10.1016/j.ijmecsci.2024.109388.

[89] Chen Y, Rui X, Zhang Z, Shehzad A. Improved incremental transfer matrix method for nonlinear rotor-bearing system. Acta Mech Sin 2020;36:1119–32. https://doi.org/10.1007/s10409-020-00976-x.

[90] Chang Z, Hou L, Lin R, Jin Y, Chen Y. A modified IHB method for nonlinear dynamic and thermal coupling analysis of rotor-bearing systems. Mechanical Systems and Signal Processing 2023;200:110586. https://doi.org/10.1016/j.ymssp.2023.110586.

[91] Yao H, Zhao Q, Xu Q, Wen B. Comparison of Common Methods in Dynamic Response Predictions of Rotor Systems with Malfunctions. International Journal of Rotating Machinery 2014;2014:1–8. https://doi.org/10.1155/2014/126843.

[92] Zhao Q, Liu J, Yuan J, Jiang H, Gao L, Zhu J, et al. Dynamic Response Analysis of Dual-Rotor System with Rubbing Fault by Dimension Reduction Incremental Harmonic Balance Method. Int J Str Stab Dyn 2022;22:2250150. https://doi.org/10.1142/S0219455422501504.

[93] Chen Y, Hou L, Lin R, Song J, Ng TY, Chen Y. A harmonic balance method combined with dimension reduction and FFT for nonlinear dynamic simulation. Mechanical Systems and Signal Processing 2024;221:111758. https://doi.org/10.1016/j.ymssp.2024.111758.

[94] Maple RC, King PI, Orkwis PD, Mitch Wolff J. Adaptive harmonic balance method for nonlinear time-periodic flows. Journal of Computational Physics 2004;193:620–41. https://doi.org/10.1016/j.jcp.2003.08.013.

[95] Jaumouillé V, Sinou JJ, Petitjean B. An adaptive harmonic balance method for predicting the nonlinear dynamic responses of mechanical systems—Application to bolted structures. Journal of Sound and Vibration 2010;329:4048–67. https://doi.org/10.1016/j.jsv.2010.04.008.

[96] Lin R, Hou L, Chen Y, Jin Y, Saeed NA, Chen Y. A novel adaptive harmonic balance method with an asymptotic harmonic selection. Appl Math Mech-Engl Ed 2023;44:1887–910.

https://doi.org/10.1007/s10483-023-3047-6.
[97] Ansys Help n.d. https://ansyshelp.ansys.com/public/account/secured?returnurl=/Views/Secured/main_page.html (accessed August 6, 2025).
[98] Krack M. maltekrack/NLvib 2025. https://github.com/maltekrack/NLvib (accessed July 25, 2025).
[99] Krack M, Gross J. Harmonic Balance for Nonlinear Vibration Problems. Cham: Springer International Publishing; 2019. https://doi.org/10.1007/978-3-030-14023-6.
[100] Krack M, Gross J. NLvib: A Matlab Tool for Nonlinear Vibration Problems. Cham: Springer International Publishing; 2019. https://doi.org/10.1007/978-3-030-14023-6.
[101] Chen H, Chen Y, Hou L, Li Z. Bifurcation analysis of rotor–squeeze film damper system with fluid inertia. Mechanism and Machine Theory 2014;81:129–39. https://doi.org/10.1016/j.mechmachtheory.2014.07.002.